\documentclass[11pt]{article}
\usepackage{latexsym}
\usepackage{amsthm}
\usepackage{amsmath}
\usepackage{graphicx}
\usepackage{amssymb}
\usepackage{color}
\usepackage{hyperref}
\usepackage{caption}
\usepackage{subcaption}
\usepackage{extarrows,mathtools,setspace,bbold}
\usepackage{longtable}
\usepackage{bm}
\usepackage{framed}
\usepackage{enumerate}
\usepackage[utf8]{inputenc}
\usepackage{tikz}
\usetikzlibrary{decorations.pathmorphing}
\usetikzlibrary{arrows}

\usepackage{colortbl}
\definecolor{summersky}{cmyk}{0.71,0.33,0,0.7}
\definecolor{flamingo}{cmyk}{0,0.51,0.71,0.6}
\definecolor{rp}{cmyk}{0.2, 1, 0.6, 0}
\definecolor{pacificblue}{cmyk}{0.95,0.3,0, 0.5}
\definecolor{gray60}{cmyk}{0.4,0.4,0,0.8}
\hypersetup{
    pdftoolbar=true,        % show AcrobatÕs toolbar?
    pdfmenubar=true,        % show AcrobatÕs menu?
    pdffitwindow=true,     % window fit to page when opened
    pdfstartview={FitH},    % fits the width of the page to the window
    pdfnewwindow=true,      % links in new window
    colorlinks=true,       % false: boxed links; true: colored links
    linkcolor=pacificblue,          % color of internal links (change box color with linkbordercolor)
    citecolor=flamingo,        % color of links to bibliography
    filecolor=magenta,      % color of file links
    urlcolor=pacificblue           % color of external links
}

\allowdisplaybreaks[3]
\newcommand\BoldSquare{%
  \setlength\fboxrule{1.1pt}\setlength\fboxsep{0pt}\fbox{\phantom{\rule{5pt}{5pt}}}}

\newcommand{\gsim}{\lower.7ex\hbox{$\;\stackrel{\textstyle>}{\sim}\;$}}
\newcommand{\lsim}{\lower.7ex\hbox{$\;\stackrel{\textstyle<}{\sim}\;$}}

\newcommand{\be}{\begin{equation}}
\newcommand{\ee}{\end{equation}}
\newcommand{\ba}{\begin{eqnarray}}
\newcommand{\ea}{\end{eqnarray}}

\newcommand{\comment}[1]{}
\newcommand{\expect}[1]{\left\langle #1 \right\rangle}
\newcommand{\mal}{\mathcal}
\newcommand{\bsb}{\boldsymbol}
\newcommand{\K}{\k}
\newcommand{\Q}{\q}

\def\d{\partial}
\def\x{\bsb x}
\def\y{{\bsb y}}
\def\xfl{\x_{\rm fl}}
\def \k {{\bsb k}}
\def\q{{\bsb q}}
\def\n{\bsb n}
\def \v{\bsb v}
\def \xfl{\x_{\rm fl}}
\def\p{{\bsb p}}
\def\z{{\bsb z}}
\def\w{{\bsb w}}
\def\tr{{\rm Tr}}
\def \unit{{\bsb 1}}
\def \O{\mathcal{O}}

\def\cH{\mal{H}}
\def\shnk{\!\!\!\!}
\def\lgr{{\rm lgr}}
\def\Pih{{\hat\Pi}}
\def\pib{{\bsb \pi}}
\def\vep{\varepsilon}
\newcommand{\intd}[1]{\int_{#1} }

\newcommand{\intq}{\int_{\bf q}}

\newcommand{\dl}{\delta}
\newcommand{\pl}{\pi}
\newcommand{\ml}{\mu}
\newcommand{\nl}{\nu}

\newcommand{\va}{\vec \alpha_{w}}
\newcommand{\vb}{\vec \beta_{w}}
\newcommand{\mb}{ \beta^{ww}}
\newcommand{\mg}{ \gamma^{w}}
\newcommand{\vg}{ \gamma^{ww}}
\newcommand{\nub}{ \bsb \nu}
\def\non{\nonumber}

\numberwithin{equation}{section}
\begin{document}

%%%%%%%%%%%%%
\vspace*{-1. cm}

\begin{center}

{\LARGE\bf Systematic Renormalization of the \\[10pt] Effective Theory of Large Scale Structure}
%{\Large\bf The Effective Theory of \\[10pt] Large Scale Structure to all Orders}
\\[1.5 cm]
{\large Ali Akbar Abolhasani,$^{a,b}$} {\large Mehrdad Mirbabayi,$^{c}$} and {\large Enrico Pajer$^{d}$}
\\[0.7cm]
%\vspace{.7cm}
%\vspace{.3cm}

{\normalsize { \sl $^a$ Physics Department, Sharif University of Technology, Tehran, Iran}}\\[10pt]
{\normalsize { \sl $^b$School of Physics, Institute for Research in Fundamental Sciences (IPM), Tehran, Iran}}\\[10pt]
{\normalsize { \sl $^c$ School of Natural Sciences, Institute for Advanced Study, Princeton, NJ 08540, USA}}\\[10pt]
{\normalsize { \sl $^d$ Institute for Theoretical Physics and Center for Extreme Matter and Emergent Phenomena, Utrecht University, Leuvenlaan 4, 3584 CE Utrecht, The Netherlands}}\\[10pt]

\vspace{.2cm}

\end{center}

\vspace{.8cm}

\hrule \vspace{0.3cm}
{  \noindent \textbf{Abstract} \\[0.3cm]
\noindent  A perturbative description of Large Scale Structure is a cornerstone of our understanding of the observed distribution of matter in the universe. Renormalization is an essential and defining step to make this description physical and predictive. Here we introduce a systematic renormalization procedure, which neatly associates counterterms to the UV-sensitive diagrams order by order, as it is commonly done in quantum field theory. As a concrete example, we renormalize the one-loop power spectrum and bispectrum of both density and velocity. In addition, we present a series of results that are valid to all orders in perturbation theory. First, we show that while systematic renormalization requires temporally non-local counterterms, in practice one can use an equivalent basis made of local operators. We give an explicit prescription to generate all counterterms allowed by the symmetries. Second, we present a formal proof of the well-known general argument that the contribution of short distance perturbations to large scale density contrast $\delta$ and momentum density $\pib(\k)$ scale as $k^2$ and $k$, respectively. Third, we demonstrate that the common practice of introducing counterterms only in the Euler equation when one is interested in correlators of $ \delta$ is indeed valid to all orders. }
 \vspace{0.3cm}
\hrule

\newpage

\tableofcontents

\newpage

%%%%%%%%%%%%%%%%%%%%
%%%%%%%%%%%%%%%%%%%%%%%%%%%%%%%%%%%%%%%%%%%%%%%%
\section{Introduction} 

A robust and accurate understanding of the gravitational clustering of Dark Matter is one of the main goals of cosmology. The smallness of the smoothed density and velocity on large scales suggests that some appropriate perturbation theory should converge to the right answer in this regime. It has recently become clear \cite{Baumann:2010tm,Carrasco:2012cv,Pajer:2013jj} that a consistent treatment of this problem requires the addition of an infinite hierarchy of effective interactions to the fluid equations of Standard Perturbation Theory (SPT) \cite{Bernardeau:2001qr}.\footnote{The discussion has by now been extended beyond dark matter \cite{Assassi:2014fva,Senatore:2014eva,Senatore:2014vja,bias,Angulo:2015eqa}. For a recent discussion of the range of validity of the effective theory see \cite{Baldauf_reach1,Baldauf_reach2,Foreman}. See also \cite{Pietroni:2011iz,Manzotti:2014loa} and \cite{Carroll,Porto:2013qua} for a different but related approach.} Although the lowest order effective interactions have been discussed in details (in the density \cite{Carrasco:2012cv,Pajer:2013jj,Carrasco:2013sva,2loop,Baldauf:2015aha} and velocity \cite{Mercolli:2013bsa} power spectrum, and in the density bispectrum \cite{Baldauf,Angulo:2014tfa}, including non-Gaussian initial conditions \cite{Assassi:2015jqa}), a complete prescription to generate all the relevant operators to all orders in perturbation theory is still missing. In addition, the current renormalization procedure -- a procedure that makes the perturbative expansion physically meaningful -- has been carried out only at the lowest order for a limited set of interesting observables, and not in a fully systematic way.

In this paper, we introduce a renormalization procedure that works systematically to all orders in perturbation theory and give an explicit prescription to construct all effective counterterms allowed by the symmetries. Here we summarize our main findings:
\begin{itemize}
\item \textbf{Systematic renormalization} In SPT one solves the continuity and Euler equations order by order:
\be
\begin{split}
\dot \delta +\theta = &-\d_i(\delta v^i)\,,\\
\dot\theta+\cH\theta+\frac{3}{2} \cH^2\delta = &-\d_i(v^j\d_j v^i).\nonumber
\end{split}
\ee
The nonlinear terms relate short and long wavelength modes to each other and lead to ``loop'' diagrams in the perturbative calculation of correlation functions. To be meaningful, loops need to be regularized and new interactions or ``counterterms'' must be included. With the exception of \cite{Mercolli:2013bsa}, all treatments so far have considered counterterms only in the Euler equation. Moreover, the SPT kernels (and the associated diagrams) combine several qualitatively different contributions together, making it practically impossible to separate embeddings of lower order loop diagrams into a higher order diagram. Because of this, the connection among various terms in the loop expansion, e.g. 1-loop 2-point correlator and 1-loop 3-point correlator, is less clear than in the common treatment of for example quantum field theories. Here, we introduce a systematic renormalization that avoid these drawbacks and is analogous to the one commonly adopted in quantum filed theories (including the Feynman diagramatics). For additional clarity, the procedure is illustrated using the explicit examples of 1-loop power spectrum and 1-loop bispectrum. 

Our systematic renormalization makes explicit the implications of the long memory of the short-scale modes, of the equivalence principle and neatly organizes the generation of new counterterms. Nevertheless, it has its own drawbacks. First, it requires the introduction of counterterms in the continuity equation. These are indeed necessary to discuss velocity correlators, but they are degenerate with those in the Euler equation if one is only interested in correlators of $ \delta $. Second, systematic renormalization requires that all counterterms are non-local in time. A practical remedy to these shortcomings is presented in section \ref{sec:remedy}.
 
\item \textbf{Mass-weighted velocity/momentum description} To renormalize the one and two-loop power spectrum and one-loop bispectrum, it is sufficient to include counterterms to the Euler equation \cite{Carrasco:2012cv,Mercolli:2013bsa,2loop,Baldauf,Angulo:2014tfa}. It is well-known that this approach works to all orders and for all statistics of $\delta$, and is equivalent to a particular definition of velocity in terms of momentum current.\footnote{It should be stressed that care is needed when comparing the prediction for velocity correlators with simulation or observations. We refer the reader to \cite{Mercolli:2013bsa} for a discussion of this issue.}  In section \ref{velind} we summarize and sharpen those arguments.

\item \textbf{All-order counterterms} We provide an explicit prescription to generate the counterterms allowed by symmetries to all orders. The discussion separates into the momentum description and the systematic renormalization approach. The latter case is more cumbersome, so we refer the interested reader directly to section \ref{list}. The result in the former case can instead be stated concisely. The Euler equation must be supplemented by 
\ba\label{finalct}
\frac{1}{1+\delta}\d_{j}\tau^{ij}\,,
\nonumber
\ea
where the two-index symmetric tensor $ \tau^{ij} $ is constructed order by order by traces and products of locally observable $ n $-th order, two-index tensors $ \Pih^{(n)}_{ij} $ and their spatial derivatives. These tensors (also called the ``Eulerian'' basis) are defined recursively by \cite{bias}
\be
\hat\Pi^{(1)}_{ij}(\x,\tau) =\: \d_i\d_j\phi(\x,\tau),\,\quad \mathrm{and} \quad \hat\Pi_{ij}^{(n)} =\:\frac{1}{(n-1)!} \left[(\hat\Pi_{ij}^{(n-1)})' - (n-1)\hat\Pi_{ij}^{(n-1)}\right],\nonumber
\ee
where prime denotes the convective derivative \eqref{convective} with respect to $\log D$ ($D$ is the linear growth factor). For example, at lowest order in derivatives and using matrix notation, the first few terms are
\ba
{\rm 1^{st}} \ && \ \unit {\rm Tr}[\Pih^{(1)}],\Pih^{(1)}  \label{listP} \nonumber \\ 
{\rm 2^{nd}} \ && \ \unit {\rm Tr}[(\Pih^{(1)})^2],\  \unit ({\rm Tr}[\Pih^{(1)}])^2,\ \Pih^{(1)} \tr[\Pih^{(1)}],\
(\Pih^{(1)})^2,\ \Pih^{(2)} \nonumber\\ 
{\rm 3^{rd}} \ && \ \unit {\rm Tr}[(\Pih^{(1)})^3 ],\ \unit {\rm Tr}[(\Pih^{(1)})^2 ]  {\rm Tr}[\Pih^{(1)}],\ \unit ({\rm Tr}[\Pih^{(1)}])^3,\ \unit {\rm Tr}[\Pih^{(1)} \Pih^{(2)}],\cdots \nonumber 
\ea
Two comments are in order. First, starting from quartic order, these operators can \textit{not} be written as local functions of (second derivatives of) the Newtonian and velocity potentials $ \phi $ and $ \phi_{v} $ \cite{bias}. Second, after taking the two derivatives of $ \tau^{ij} $ appearing in \eqref{finalct}, some terms become degenerate. For example, there are five second order terms above but only 3 independent counterterms at this order \cite{Baldauf,Angulo:2014tfa}.

\item \textbf{Double softness} It is well-known (see e.g. \cite{Peebles}) that the conservation of momentum and locality implies that interactions among modes of characteristic wave-number $q$ can generate a longer wavelength mode $k\ll q$ suppressed at least as $ \delta(k)\propto k^{2} $, namely the density is ``double soft''. Analogously, the momentum is single soft, $ |\pib|\propto k $. After reviewing a general argument for this property due to Peebles, we prove it directly from the equations of motion in section \ref{double} (details are in appendix \ref{app:soft}). Double softness can be seen and implemented most straghtforwardly using mass-weighted velocity or momentum. On the other hand, in terms of the more general velocity used in \cite{Mercolli:2013bsa} and in the systematic renormalization, double softness implies a highly non-trivial relation among counterterms to all orders.

\end{itemize}

In this paper, we have extensively used the same nomenclature as in quantum and statistical field theory to stress the similarity with the problem at hand and to make the discussion more intuitive. When needed, we have adapted and refined some of the definitions. For the convenience of the reader, a glossary of the technical terms is provided in appendix \ref{app:glos}.

The rest of the paper is organized as follows. In section \ref{pre}, we review standard perturbation theory in a compact ``vector'' form and introduce diagrammatic rules, which we elucidate for the case of the tree-level bispectrum. In section \ref{loops}, we discuss how interactions give rise to loops and the separation between 1-particle irreducible (1PI) and 1-particle reducible (1PR) diagrams and stochastic and non-stochastic counterterms. The one-loop power spectrum is used as an example. In section \ref{counter}, we present our systematic renormalization procedure and discuss the implication of the long memory of the short-scale modes and of the equivalence principle. The one-loop bispectrum is used as a concrete example of our general remarks. Section \ref{double} is devoted to the discussion of double softness. Finally, in section \ref{sec:remedy}, we give the explicit form of all allowed counterterms and put the density-only description on a firmer footing. Some technical details are left to the appendices.

%%%%%%%%%%%%%%%%%%%%%%%%%%%%%%%%%%%%%%%%%%%%%%%%%
\section{Preliminaries}\label{pre}

At very large scales dark matter can be treated effectively as a pressure-less fluid of density contrast $\delta = (\rho-\bar\rho)/\bar\rho$, and velocity $v^i$, whose dynamics (in a matter dominated model of expansion rate $\cH$) are governed by the continuity and Euler equations:
\be
\begin{split}\label{sptcoupled}
\dot \delta +\theta = &-\d_i(\delta v^i)\,,\\
\dot \v + \cH \v +\nabla \phi = & -\v\cdot \nabla \v,
\end{split}
\ee
where the gravitational potential satisfies $\nabla^2 \phi = \frac{3}{2} \cH^2\delta$. Starting from zero vorticity, i.e. $\nabla \times \v =0$, it remains zero under this evolution. Hence, we can take the divergence of the latter equation and use the velocity divergence as the independent variable $\theta=\d_i v^i$:
\be\label{thetaeq}
\dot \theta+ \cH \theta +\frac{3}{2}\cH^2 \delta =  -\d_i(v^j\d_j v^i).
\ee
(We will comment on the generation of vorticity in section \ref{sec:soft}.) At shorter scales the above description is incomplete and higher derivative corrections must be included. In practice, the need for such corrections (counterterms) can be seen from the dependence of the nonlinear solutions on the choice of the ultraviolet cutoff. The coefficients of the counterterms depend on the cutoff in such a way that the final result is cutoff-independent, though it does depend on the unknown details of the ultraviolet physics. That is, the counterterms capture the contribution of short distance physics by introducing new effective interactions among long-wavelength modes. Their coefficients are to be fixed either by a first-principle calculation or, more realistically, using simulations and observations. In this paper, we will derive the general form of these counterterms.  

For this purpose, it is useful to develop a systematic approach to renormalization---the procedure of removing cutoff dependence---as in conventional relativistic field theories. The main difference is that here we are dealing with an initial value problem, whose implications becomes clear in the sequel. In particular, it helps to have a more symmetric treatment of the primary variables $\delta$ and $\theta$, and a more explicit diagrammatic representation of perturbation theory. Throughout the paper we work in an Einstein-de Sitter (EdS) universe but to a good approximation the conclusions remain unchagned in $ \Lambda $CDM cosmology. Following \cite{Crocce:2005xy,Bernardeau:2013oda} we introduce a doublet field
\ba
\Psi_a  = \left(\delta,-\dfrac{1}{\cal H} \theta \right)\,.
\ea
The equations of motion can then be written (in momentum space) as
\ba
\label{main-eq}
\dfrac{\partial}{\partial \eta} \Psi_a (\bsb{k}, \eta)+ \Omega_{ab} (\eta) \Psi_b(\bsb{k}, \eta) =\int_{\k_1,\k_2} 
\gamma^{(s)}_{abc}(\K,\K_1,\K_2)\,\Psi_b(\bsb{k}_1, \eta) \Psi_c(\bsb{k}_2, \eta),
\ea
where
\ba
\int_\k &\equiv &\int  \frac{d^3 \k}{(2\pi)^3} \,,\quad  \eta = \log a\,, \quad\Omega_{a\,b} = 
 \Bigg[  \begin{array}{ccc}
0 & -1\\
-3/2 & 1/2
\end{array}
\Bigg]\,,
\ea
and non-vanishing  components of the symmetrized vertex functions $\gamma^{(s)}$ are given by 
\ba\label{vert}
\gamma_{121} (\bsb{k},\bsb{k_1},\bsb{k_2}) = \delta^3 (\bsb{k}-\bsb{k_1}-\bsb{k_2}) \dfrac{\alpha(\bsb{k_1},\bsb{k_2})}{2}\,,
\\
\gamma_{112} (\bsb{k},\bsb{k_1},\bsb{k_2}) = \delta^3 (\bsb{k}-\bsb{k_1}-\bsb{k_2}) \dfrac{\alpha(\bsb{k_2},\bsb{k_1})}{2}\,,
\\
\gamma_{222} (\bsb{k},\bsb{k_1},\bsb{k_2}) = \delta^3 (\bsb{k}-\bsb{k_1}-\bsb{k_2}) \beta(\bsb{k_1},\bsb{k_2})\,.
\ea
The vertices can be read from \eqref{sptcoupled}:
\be\label{ab}
\alpha(\k_1,\k_2)= \frac{\k_1\cdot (\k_1+\k_2)}{k_1^2}\,,\quad 
\beta(\k_1,\k_2)= \frac{\k_1\cdot\k_2 \:(\k_1+\k_2)^2}{2\, k_1^2 k_2^2}\,.
\ee

%%%%%%%%%%%%%%%%%%%%%%%%%%%%%%%%%%
\subsection{Perturbation theory and diagrammatic rules}

Given some initial condition $\Psi_a(\k,\eta_0)=(\delta_1(\k),\delta_1(\k))$, the above system can be solved perturbatively to find $\Psi_a(\k,\tau)$. At first order, we have
\be
\Psi^{(1)}_a(\k,\eta)=D^+_{ab}(\eta,\eta_0)\Psi_{b}(\k,\eta_0),
\ee
where repeated indices are summed over and $D^+_{ab}$ is the growing solution of the linearized system. Note that it is $\k$-independent. $D^+_{ab}(\eta)$ is related to the conventional growth factor through
\be
D(\eta)= \sum_a D_{1 a}^+(\eta,\eta_0)= \sum_a D_{2 a}^+(\eta,\eta_0).
\ee
The higher order solutions can be expressed, recursively, by combining two lower order ones via the interactions vertices $\gamma_{abc}$:
\be\label{psin}
\Psi^{(n)}_a (\K, \eta) = \sum_{m=1}^{n-1} \int_{\k_1,\k_2}\int_{\eta_0}^{\eta} d \eta' g_{a\,b} (\eta-\eta') 
\gamma_{b\,c\,d} (\K,\K_1,\K_2,\eta') \Psi^{(m)}_c(\K_1,\eta') \Psi^{(n-m)}_d(\K_2,\eta'),
\ee
where the retarded Green's function, or \textit{propagator}, is given by
\ba
\label{propagator}
g_{ab}(\eta ,\eta ')= \dfrac{e^{\eta - \eta '}}{5} \Bigg[  \begin{array}{cc}
3 & 2
\\
3 & 2
\end{array}
\Bigg]
+\dfrac{e^{-\frac{3}{2}(\eta-\eta')}}{5} \Bigg[  \begin{array}{cc}
-2 & 2
\\
3 & -3
\end{array}
\Bigg]\,,
\ea
for $\eta>\eta'$, and $0$ otherwise. The growth function $D^+_{ab}(\eta,\eta_0)$ coincides with the growing part of $g_{ab}(\eta,\eta_0)$ and, since $\eta_0$ is practically $-\infty$, it can be approximated by $g_{ab}(\eta,\eta_0)$. This perturbative expansion is diagrammatically illustrated by Feynman diagrams as
\be\label{Feyn}
\raisebox{-15 pt}{
\begin{tikzpicture}[scale=.7]
\draw [very  thick] (0,0) -- (4,0);
\draw (2,0) node {$\blacktriangleleft$};
\draw (0,-.5) node {$a,\eta$};
\draw (4,-.5) node {$b,\eta'$};
\draw (2,.5) node {$\K$};
\draw (-2,0) node {$g_{a\,b}(\eta,\eta'):$};
\end{tikzpicture}
}\quad
\raisebox{-37 pt}{
\begin{tikzpicture}[scale=.7]
\draw [very  thick](0,0) -- (2,0);
\draw [very  thick](2,0) -- (4,1.5);
\draw [very  thick](2,0) -- (4,-1.5);
\draw (3,-.75) node [rotate=-35]{$\blacktriangleleft$};
\draw (3,.75) node [rotate=35]{$\blacktriangleleft$};
\draw (1,0) node {$\blacktriangleleft$};
%\draw (1,.5) node {$\K$};
%\draw (3,1.25) node {$\K_1$};
%\draw (3,-1.25) node {$\K_2$};
\draw (-.7,0) node {$a, \K$};
\draw (4.5,1.8) node {$a_1,\K_1$};
\draw (4.5,-1.8) node {$a_2,\K_2$};
\draw (-3.75,0) node {$\gamma_{a\,a_1\,a_2}(\K,\K_1,\K_2):$};
\end{tikzpicture}
}
\ee
Every propagator is represented by a line, and the cubic interactions by vertices. The arrows indicate the flow of time, and clearly the first index of $\gamma_{abc}$, which corresponds to the out-going line, has no symmetry property with the last two which refer to the in-going lines. A typical diagram contributing to $\Psi^{(4)}$ is 
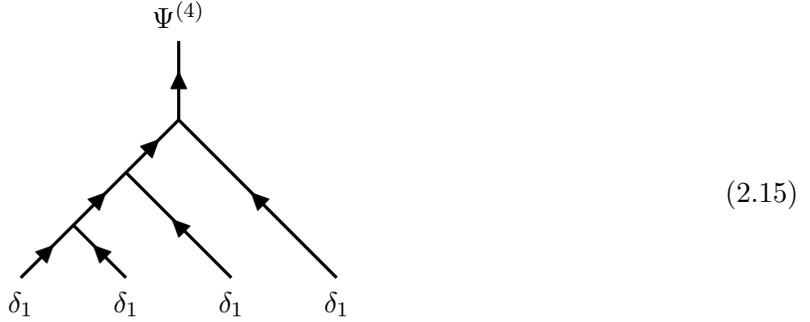
\be\label{Psi4}
\raisebox{-47 pt}{
\begin{tikzpicture}[scale=.7]
\draw [very  thick](0,0) -- (3,3);
\draw [very  thick](1,1) -- (2,0);
\draw [very  thick](2,2) -- (4,0);
\draw [very  thick](3,3) -- (6,0);
\draw [very  thick](3,3) -- (3,4.5);

\draw (.5,.5) node [rotate=45]{$\blacktriangleright$};
\draw (1.5,1.5) node [rotate=45] {$\blacktriangleright$};
\draw (2.5,2.5) node [rotate=45] {$\blacktriangleright$};
\draw (3,3.75) node [rotate=90] {$\blacktriangleright$};
\draw (1.5,.5) node [rotate=-45]{$\blacktriangleleft$};
\draw (2.5,2.5) node [rotate=45] {$\blacktriangleright$};
\draw (3,1) node [rotate=-45] {$\blacktriangleleft$};
\draw (4.5,1.5) node [rotate=-45] {$\blacktriangleleft$};

\draw (0,-.5) node  {$\delta_1$};
\draw (2,-.5) node  {$\delta_1$};
\draw (4,-.5) node  {$\delta_1$};
\draw (6,-.5) node  {$\delta_1$};
\draw (3,5) node  {$\Psi^{(4)}$};

\end{tikzpicture}
}
\ee
where here and in what follows the indices are often omitted to avoid clutter. Note also that the momentum integrals in \eqref{psin} must be understood as running up to some ultraviolet cutoff $\Lambda$ beyond which the effective theory is not applicable anymore. Renormalization ensures that the final results do not depend on the choice of $\Lambda$.

Using the above Feynman rules, the following recipe can be used to calculate correlation functions perturbatively. An external leg $\Psi_a^{(n)}$ is given by the sum over all tree diagrams constructed by successive bifurcations through trilinear interaction vertices to get to $n$ initial fields $\delta_1(\k)$, as in \eqref{Psi4}. Any trilinear vertex merges a $\Psi^{(n)}$ and a $\Psi^{(m)}$ to result in an $(n+m)$-th order $\Psi^{(n+m)}$ field. These perturbative expansions for the external fields are then correlated using the initial statistics of $\delta_1(\k)$. We assume Gaussian initial condition
\be
\expect{\delta_1(\k)\delta_1(\k')}= P_{\rm lin}(k)(2\pi)^3 \delta^3(\k+\k'),
\ee
which reduces the correlation functions of several $\Psi$ fields to a sum over all possible pairings of the initial fields that appear in their perturbative expansion. The correlation of initial fields is called \textit{contraction} and is reperesented by a dot. If all initial momenta in a given diagram are fixed in terms of the external ones, the diagram is called tree-level. One example is
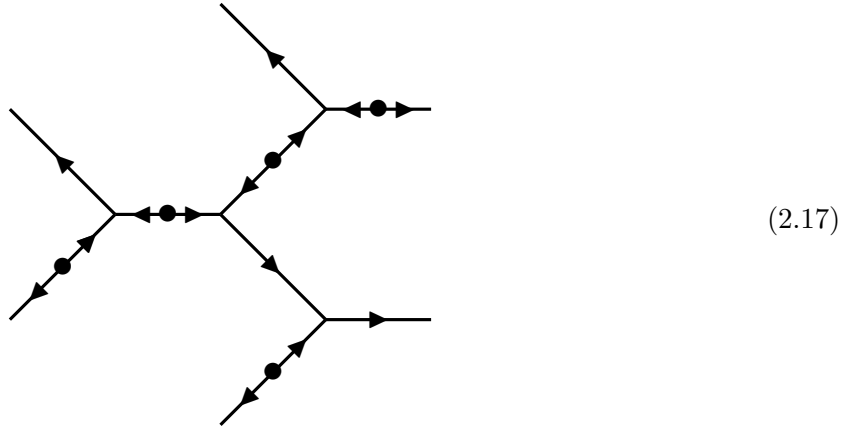
\be\label{tree}
\raisebox{-75 pt}{
\begin{tikzpicture}[scale=.7]
\draw [very  thick](0,-2) -- (2,0);
\draw [very  thick](0,2) -- (2,0);
\draw [very  thick](2,0) -- (4,0);
\draw [very  thick](4,0) -- (6,2);
\draw [very  thick](4,0) -- (6,-2);
\draw [very  thick](6,2) -- (4,4);
\draw [very  thick](6,-2) -- (4,-4);
\draw [very  thick](6,2) -- (8,2);
\draw [very  thick](6,-2) -- (8,-2);

\draw (1,-1) node {\scalebox{1.5}{$\bullet$}};
\draw (3,0) node {\scalebox{1.5}{$\bullet$}};
\draw (5,1) node {\scalebox{1.5}{$\bullet$}};
\draw (7,2) node  {\scalebox{1.5}{$\bullet$}};
\draw (5,-3) node  {\scalebox{1.5}{$\bullet$}};

\draw (3.5,0) node {$\blacktriangleright$};
\draw (2.5,0) node {$\blacktriangleleft$};
\draw (6.5,2) node {$\blacktriangleleft$};
\draw (7.5,2) node  {$\blacktriangleright$};
\draw (7,-2) node {$\blacktriangleright$};

\draw (1.5,-.5) node [rotate=45]{$\blacktriangleright$};
\draw (.5,-1.5) node [rotate=45] {$\blacktriangleleft$};
\draw (1,1) node [rotate=-45] {$\blacktriangleleft$};
\draw (1.5,-.5) node [rotate=45]{$\blacktriangleright$};
\draw (4.5,.5) node [rotate=45] {$\blacktriangleleft$};
\draw (5.5,1.5) node [rotate=45] {$\blacktriangleright$};
\draw (5,-1) node [rotate=-45] {$\blacktriangleright$};
\draw (4.5,-3.5) node [rotate=45] {$\blacktriangleleft$};
\draw (5.5,-2.5) node [rotate=45] {$\blacktriangleright$};
\draw (5,3) node [rotate=-45] {$\blacktriangleleft$};
\end{tikzpicture}
}
\ee
Otherwise, if there are undetermined initial momenta it is a loop diagram. There is one undetermined momentum for each loop and it must be integrated over. The tree-level contribution to an $N$-point correlation function is of order $2(N-1)$ in $\delta_1$, and an $n$-loop contribution is of order $2(N+n-1)$. It is common to label various diagrams by the orders of their external legs in perturbation theory. For instance, the 1-loop contribution to the power spectrum where $\Psi^{(1)}$ is correlated with $\Psi^{(3)}$ is often called $P_{13}$. 

%%%%%%%%%%%%%%%%%%%%%%%%%%%%%%%
\subsection{Example: tree-level bispectrum}

Let us conclude by illustrating the above formalism in the example of tree-level 3-point correlation function of density contrast 
\be
\expect{\delta(\k_1,\eta)\delta(\k_2,\eta)\delta(\k_3,\eta)}=(2\pi)^3\delta^3(\k_1+\k_2+\k_3)B^{\delta}_{112}(\k_1,\k_2,\k_3,\eta).
\ee
In the conventional SPT formalism this is given in terms of the $F_n$ kernels:
\be\label{B112}
B^{\delta}_{112} (\K_1,\K_2,\K_3,\eta) =2e^{4\eta}F_2(\K_1,\K_2) P_{\mathrm{lin}}(k_1) P_{\mathrm{lin}}(k_2) + 2 \mathrm{cycl. perm.}\,,
\ee
where
\be
F_2(\k_1,\k_2)=\frac{5}{7}+\frac{2}{7}\frac{(\k_1\cdot\k_2)^2}{k_1^2k_2^2}
+\frac{1}{2}\k_1\cdot\k_2(\frac{1}{k_1^2}+\frac{1}{k_2^2}).
\ee
In the doublet formalism, the relevant diagram is 
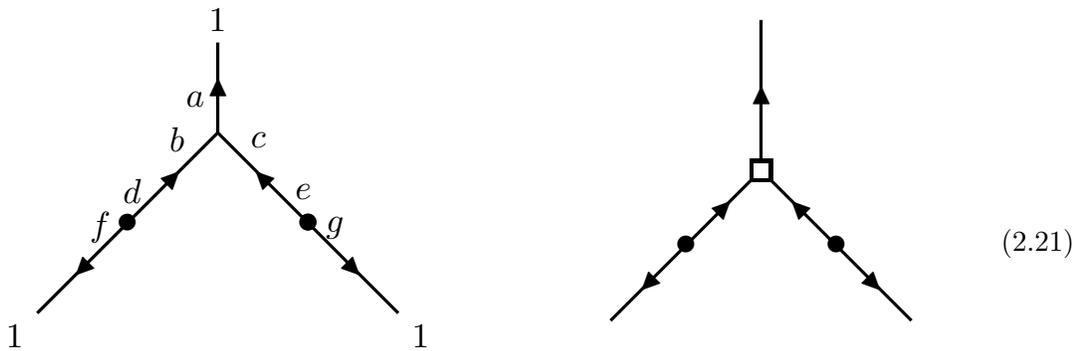
\be\label{SPT_doublet_B}
\raisebox{-40 pt}{
\begin{tikzpicture}[scale=.6]

\draw [very  thick](0,0) -- (0,2);
\draw [very thick](-4,-4) -- (0,0);
\draw [very  thick](4,-4) -- (0,0);

\draw (-2,-2) node {\scalebox{1.5}{$\bullet$}};
\draw (2,-2) node {\scalebox{1.5}{$\bullet$}};

\draw (0,1) node [rotate=90]{$\blacktriangleright$};
\draw (-1,-1) node [rotate=45]{$\blacktriangleright$};
\draw (-3,-3) node [rotate=45] {$\blacktriangleleft$};
\draw (1,-1) node [rotate=-45] {$\blacktriangleleft$};
\draw (3,-3) node [rotate=-45]{$\blacktriangleright$};

\draw (0,2.5) node {\scalebox{1.2}{${1}$}};
\draw (-4.5,-4.5) node {\scalebox{1.2}{${1}$}};
\draw (4.5,-4.5) node {\scalebox{1.2}{${1}$}};
\draw (-.5,.75) node {\scalebox{1.2}{${a}$}};
\draw (-.9,-.15) node {\scalebox{1.2}{${b}$}};
\draw (.9,-.15) node {\scalebox{1.2}{${c}$}};
\draw (-1.9,-1.3) node {\scalebox{1.2}{${d}$}};
\draw (1.9,-1.3) node {\scalebox{1.2}{${e}$}};
\draw (-2.6,-2.1) node {\scalebox{1.2}{${f}$}};
\draw (2.6,-2.1) node {\scalebox{1.2}{${g}$}};

\draw (2,-2) node {\scalebox{1.5}{$\bullet$}};

\end{tikzpicture}
}
\hspace{2 cm}
\raisebox{-27 pt}{
\begin{tikzpicture}[scale=1]
\draw [very  thick](0,.15) -- (0,2);
\draw [very  thick](-2,-2) -- (-.13,-.13);
\draw [very  thick](2,-2) -- (.13,-.13);

\draw (-1,-1) node {\scalebox{1.5}{$\bullet$}};
\draw (1,-1) node {\scalebox{1.5}{$\bullet$}};
\draw (0,0) node {\scalebox{1.2}{$\BoldSquare$}};

\draw (0,1) node[rotate=90] {$\blacktriangleright$};

\draw (-.5,-.5) node [rotate=45]{$\blacktriangleright$};
\draw (-1.5,-1.5) node [rotate=45] {$\blacktriangleleft$};
\draw (.5,-.5) node [rotate=-45] {$\blacktriangleleft$};
\draw (1.5,-1.5) node [rotate=-45]{$\blacktriangleright$};
\end{tikzpicture}
}
\ee
where we demonstrated the SPT diagram on the right for comparison. Following the Feynman rules, one obtains 
\be
\nonumber
\sum_{a\,b\,c \, d\,e\,f\,g}\int^{\eta}_{\eta_0}\shnk d\eta'  g_{1\,a} (\eta,\eta') \,g_{b\,d} (\eta',\eta_0) 
\,g_{c\,e} (\eta',\eta_0)\,g_{1\,f} (\eta,\eta_0) \,g_{1\,g} (\eta,\eta_0) \gamma_{a\,b\,c} (\K_1,\K_2,\K_3) 
 P_{\mathrm{lin}}(k_1) P_{\mathrm{lin}}(k_2).
\ee
Using the above expressions for $g_{ab}$ and $\gamma_{abc}$ (and, of course, including other permutations), it is easy to show that this agrees with \eqref{B112}.

%%%%%%%%%%%%%%%%%%%%%%%%%%%%%%%%%%%%%%%%%%%%%%%
\section{Loops}\label{loops}

Loops are encountered in the perturbative calculation of correlators when the initial momenta are not fixed in terms of external momenta, and therefore, are integrated over. The ultraviolet part of these integrals include short wavelength modes which are not expected to be well described by the effective field theory. The integrals are often divergent. One should therefore regulate the integrals (say by sharply cutting them off at momentum $\Lambda$), and use the cutoff dependence to introduce new effective interactions (counterterms) that capture the contribution of the short scale physics to the low momentum observables as an expansion in $k/\Lambda$. This procedure provides a systematic way of inferring the form of higher derivative corrections to be added to the long-wavelength effective theory, and in most cases, it covers all possibilities that are compatible with the symmetries of the system under consideration.\footnote{There can be exceptions when some higher derivative operator is protected by a non-renormalization theorem.}

Hence, understanding better the properties of loop diagrams is the key to identify the general structure of higher derivative corrections. This is the subject of this section.

%%%%%%%%%%%%%%%%%%%%%%%%%%%%%%%%%%%%%%%%%%%%%%%%%%%%%%%%%%%%%%%%%%%%%%%%%%%%%%%%
\subsection{Reducible and irreducible diagrams}

Consider the $P_{13}$ contribution to the 2-point correlation function 
\be\label{P13}
\raisebox{-33 pt}{
\begin{tikzpicture}[scale=.8]
\draw[very thick] (0,0) -- (1,0);
\draw [very thick](3,0) -- (5,0);
\draw[-{triangle 45}] (2,.75) arc (90:35:1cm and .75cm) ;
\draw[-{triangle 45}] (2,.75) arc (90:145:1cm and .75cm) ;

\draw (3.5,0) node {$\blacktriangleleft$};
\draw (4.5,0) node {$\blacktriangleright$};
\draw (2,-.75) node {$\blacktriangleleft$};
\draw (.5,0) node {$\blacktriangleleft$};

\draw (2,1.3) node {$P(q)$};
\draw (4,.5) node {$P(k)$};
\draw (-.3,0.1) node {$\k$};
\draw (5.5,0.1) node {$-\k$};
\draw (2,-1.15) node {$\k-\q$};

\draw (2,.75) node {\scalebox{1.5}{$\bullet$}};
\draw (4,0) node {\scalebox{1.5}{$\bullet$}};
\draw [very  thick](3,0) arc (0:-180:1cm and .75cm);
\draw [very  thick](3,0) arc (0:81:1cm and .75cm);
\draw [very  thick](1,0) arc (180:98:1cm and .75cm);
\end{tikzpicture}
}
\ee
Suppressing time-integrations and indices the diagram would have the following momentum structure
\be
P_{\rm lin}(k)\int d^3\q P_{\rm lin}(q) \gamma(\k,\q,\k-\q)\gamma(\k-\q,\k,-\q).
\ee
To renormalize this loop integral a linear counterterm must be added to the equations of motion. Its leading contribution to the 2-point correlation function 
\be\label{counter-term}
\raisebox{-5 pt}{
\begin{tikzpicture}[scale=.7]
\draw [very  thick](0,0) -- (1.3,0);
\draw [very  thick](1.68,0) -- (5,0);
\draw [very  thick](1.5,0) node {\scalebox{.9}{$\bm{\otimes}$}};
\draw [very  thick](3.5,0) node {\scalebox{1.5}{$\bullet$}};

\draw (2.5,0) node {$\blacktriangleleft$};
\draw (4.5,0) node {$\blacktriangleright$};
\draw (.5,0) node {$\blacktriangleleft$};

\end{tikzpicture}
}
\ee
combines with the original diagram to make the full result cutoff-independent. 

Now suppose the same loop diagram of \eqref{P13} is inserted into a higher order graph, e.g.
\be\label{B1PR}
\raisebox{-35 pt}{
\begin{tikzpicture}[scale=.7]
\draw [very  thick](0,0) -- (1,0);
\draw [very  thick](3,0) -- (4,0);
\draw [very  thick](3,0) arc (0:-180:1cm and .75cm);
\draw [very  thick](3,0) arc (0:83:1cm and .75cm);
\draw [very  thick](1,0) arc (180:97:1cm and .75cm);
\draw [very  thick](4,0) -- (6,2);
\draw [very  thick](4,0) -- (6,-2);

\draw[-{triangle 45}] (2,.75) arc (90:35:1cm and .75cm) ;
\draw[-{triangle 45}] (2,.75) arc (90:145:1cm and .75cm) ;
\draw (3.5,0) node {$\blacktriangleleft$};
\draw (2,-.75) node {$\blacktriangleleft$};
\draw (.5,0) node {$\blacktriangleleft$};
\draw (4.5,.5) node [rotate=45] {$\blacktriangleleft$};
\draw (4.5,-.5) node [rotate=-45] {$\blacktriangleleft$};
\draw (5.5,1.5) node [rotate=45] {$\blacktriangleright$};
\draw (5.5,-1.5) node [rotate=-45] {$\blacktriangleright$};

\draw (5,1) node {\scalebox{1.5}{$\bullet$}};
\draw (5,-1) node {\scalebox{1.5}{$\bullet$}};
\draw (2,.75) node {\scalebox{1.5}{$\bullet$}};

\end{tikzpicture}
}
\ee
One would expect that if the counterterm \eqref{counter-term} is chosen appropriately, inserting the same quadratic vertex in place of the loop diagram, namely
\be\label{ctB1PR}
\raisebox{-35 pt}{
\begin{tikzpicture}[scale=.7]
\draw [very  thick](-.6,0) -- (.51,0);
\draw [very  thick](.89,0) -- (2,0);
\draw [very  thick](2,0) -- (4,2);
\draw [very  thick](2,0) -- (4,-2);

\draw (0,0) node {$\blacktriangleleft$};
\draw (1.35,0) node {$\blacktriangleleft$};
\draw (2.5,.5) node [rotate=45] {$\blacktriangleleft$};
\draw (3.5,1.5) node [rotate=45] {$\blacktriangleright$};
\draw (2.5,-.5) node [rotate=-45] {$\blacktriangleleft$};
\draw (3.5,-1.5) node [rotate=-45] {$\blacktriangleright$};

\draw (3,1) node {\scalebox{1.5}{$\bullet$}};
\draw (3,-1) node {\scalebox{1.5}{$\bullet$}};
\draw (.7,0) node {\scalebox{.9}{$\bm{\otimes}$}};

\end{tikzpicture}
}
\ee
should renormalize \eqref{B1PR}. We will see that such a choice of the counterterm is indeed possible, and in particular, the diagram \eqref{ctB1PR} does renormalize diagram \eqref{B1PR}. Hence, embedding lower order loop diagrams in higher order ones does not lead to genuinely new corrections (counterterms). We call these loop diagrams reducible, or in analogy with particle physics diagrammatics, \textit{1-Particle Reducible} (1PR). Their characteristic property is that there exist an internal line $l$ that is not immediately connected to any external line through a contraction such that cutting $l$ divides the diagram into two disconnected pieces.
% by cutting an internal line that is not immediately connected to a contraction (i.e.~a line that represents $\Psi^{(n)}$ with $n$ strictly larger than $1$ but less than the order of the external field it will lead to by following the time-arrows). 
All other diagrams are called {\em 1-Particle Irreducible} (1PI) diagrams, \eqref{P13} is an example. 

Since the counterterms are to be added to the equations of motion as new vertices, it makes sense to work not with the full diagrams representing the correlation functions but directly with the time-evolution diagrams. These show the evolution of a set of initial fields into one higher order final field at a later time. The loops correspond to integrating over some of the initial momenta, and hence, to allowing them to become ``hard'', i.e.~of the order of the ultraviolet cutoff $\Lambda$. We represent these hard momenta by thick lines. Moreover, to exclude 1PR diagrams one would simply amputate all other (``soft'') lines. That is, the relevant diagrams are made of (i) several hard lines extending all the way to the initial condition $\delta_1$, (ii) one soft out-going line, and (iii) any number of soft in-going lines which are immediately connected to the hard lines. An example is
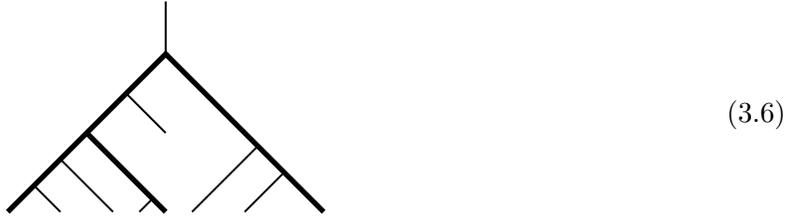
\be\label{t1PIa}
\raisebox{-35 pt}{
\begin{tikzpicture}[scale=.35]
\draw [line width= 2 pt](0,0) -- (6.07,6.07);
\draw [line width= 2 pt](3,3) -- (6,0);
\draw [line width= 2 pt](6,6) -- (12,0);
\draw [line width= .75 pt] (1,1) -- (2,0);
\draw [line width= .75 pt] (2,2) -- (4,0);
\draw [line width= .75 pt] (5,0) -- (5.5,.5);
\draw [line width= .75 pt] (7,0) -- (9.5,2.5);
\draw [line width= .75 pt] (9,0) -- (10.5,1.5);
\draw [line width= .75 pt] (6,3) -- (4.5,4.5);
\draw [line width= .75 pt] (6,6) -- (6,8);
\end{tikzpicture}
}
\ee
The presence of high-momentum hard lines makes these diagrams cutoff dependent. They are renormalized by adding new vertices to the effective theory that have the same structure of the soft lines, and the same cutoff dependence in their coefficients:
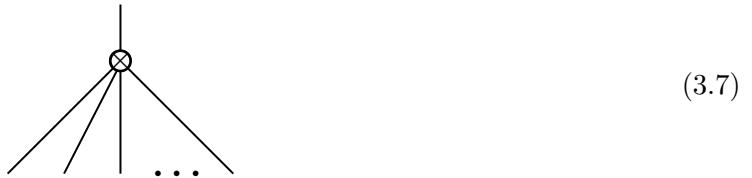
\be\label{t1PIb}
\raisebox{-35 pt}{
\begin{tikzpicture}[scale=.5]

\draw [line width= .75 pt] (3,3.25) -- (3,4.5);
\draw [line width= .75 pt] (0,0) -- (2.8,2.8);
\draw [line width= .75 pt] (6,0) -- (3.2,2.8);
\draw [line width= .75 pt] (3,0) -- (3,2.75);
\draw [line width= .75 pt] (1.5,0) -- (2.91,2.75);
\draw [line width= 1. pt] (3.27,3)  arc (0:360:.27cm and .27cm) ;

\draw (3,3) node {$\bm{\otimes}$};
%\draw (3.5,0) node {\scalebox{.5}{$\bullet$}};
\draw (4,0) node {\scalebox{.5}{$\bullet$}};
\draw (4.5,0) node {\scalebox{.5}{$\bullet$}};
\draw (5,0) node {\scalebox{.5}{$\bullet$}};

\end{tikzpicture}
}
\ee
Note that the contribution of the new vertex to the final field always comes with the propagator of the out-going soft field which is common between \eqref{t1PIa} and \eqref{t1PIb}. Since we are interested in the structure of the corrections to the equations of motion and not the resulting final field, we will also amputate the final propagator. 

%%%%%%%%%%%%%%%%%%%%%%%%%%%%%%%%%%%%%%%%%%%%%%%%%%%%%%%%%%%
\subsection{Stochastic vs. non-stochastic corrections}

There is another important classification of counterterms that is unique to the case of initial value problem. The high-momentum initial fields were seen to result from integration over unfixed initial momenta. If in a given time-evolution diagram all of these can be paired and contracted with each other, the associated counterterm is called a \textit{non-stochastic} one, e.g.
\be\label{Psi3}
\textrm{non-stochastic diagram:}\quad
\raisebox{-55 pt}
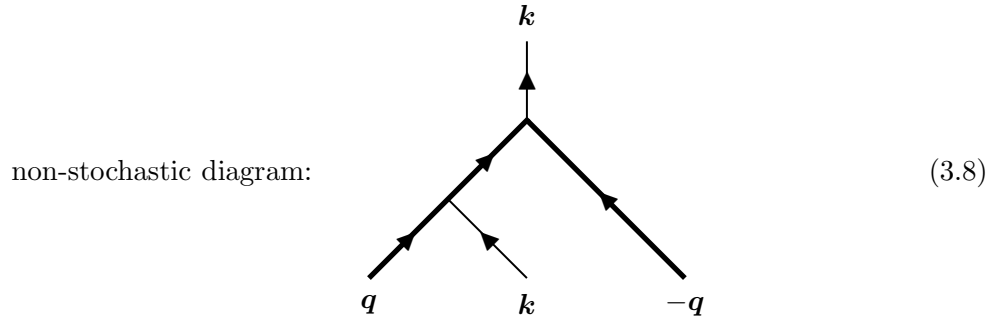
{
\begin{tikzpicture}[scale=.7]
\draw [line width= 2 pt](0,0) -- (3.035,3.035);
\draw [line width= 2 pt](3,3) -- (6,0);

\draw [line width= .75 pt](3,3) -- (3,4.5);
\draw [line width= .75 pt](3,0) -- (1.5,1.5);

\draw (.75,.75) node [rotate=45] {$\blacktriangleright$};
\draw (2.25,2.25) node [rotate=45] {$\blacktriangleright$};
\draw (2.25,.75) node [rotate=-45] {$\blacktriangleleft$};
\draw (3,3.75) node [rotate=90] {$\blacktriangleright$};
\draw (4.5,1.5) node [rotate=-45] {$\blacktriangleleft$};

\draw (0,-.5) node  {$\Q$};
\draw (3,-.5) node  {$\K$};
\draw (6,-.5) node  {$-\Q$};
\draw (3,5) node  {$\K$};

\end{tikzpicture}
}
\ee
The counterterms of these diagram capture the fact that short scales are affected by large scales. Otherwise, if at least one pair of the initial hard fields do not have exactly opposite momenta, it corresponds to the probability that short scale fluctuations average into a long-wavelength mode, e.g.
\be\label{Psi2}
\textrm{stochastic diagram:}\quad
\raisebox{-55 pt}{
\begin{tikzpicture}[scale=.7]
\draw [line width= 2 pt](0,0) -- (3.035,3.035);
\draw [line width= 2 pt](3,3) -- (6,0);

\draw [line width= .75 pt] (3,3) -- (3,4.5);

\draw (1.5,1.5) node [rotate=45] {$\blacktriangleright$};
\draw (3,3.75) node [rotate=90] {$\blacktriangleright$};
\draw (4.5,1.5) node [rotate=-45] {$\blacktriangleleft$};

\draw (0,-.5) node  {$\Q$};
\draw (6,-.5) node  {$-\Q+\K$};
\draw (3,5) node  {$\K$};

\end{tikzpicture}
}
\ee
The associated counterterms are called \textit{stochastic}. These are by nature indeterministic and hence contain a stochastic field (usually denoted by $J(x,\eta)$) about which we can only make statistical statements.

One can contract and integrate over the momenta of all initial hard fields that are paired in a time-evolution diagram (which include all of the hard lines in a non-stochastic diagram). This is because the loop is independent of how the ingoing soft lines are embedded in a more complicated diagram. For instance, the hard modes $\delta_1(\q)$ and $\delta_1(-\q)$ in fig. \ref{Psi3} can be contracted regardless of whether we consider diagram of \eqref{P13} or \eqref{B1PR}. After doing so these diagrams (and the associated counter-terms) incorporate the average response of the short modes to the initial long modes.\footnote{Notice that the vertices \eqref{vert} vanish for zero outgoing momentum since $ \alpha(\k,-\k)=\beta (\k,-\k)=0 $ (see section \ref{sec:soft} for more details). Therefore, non-stochastic diagrams with no ingoing soft lines all vanish.} On the other hand, the unpaired initial hard modes will be contracted by unpaired modes coming from other external lines as in
\be\label{P22}
\raisebox{-19 pt}{
\begin{tikzpicture}[scale=.7]

\draw[very  thick] (-1,0) -- (1,0);
\draw [very  thick] (5,0) -- (3,0);
\draw [very  thick](3,0) arc (0:-180:1cm and .75cm);
\draw [very  thick](3,0) arc (0:180:1cm and .75cm);

\draw[-{triangle 45}] (2,.75) arc (90:35:1cm and .75cm) ;
\draw[-{triangle 45}] (2,.75) arc (90:145:1cm and .75cm) ;
\draw[-{triangle 45}] (2,-.75) arc (-90:-35:1cm and .75cm) ;
\draw[-{triangle 45}] (2,-.75) arc (-90:-145:1cm and .75cm) ;

\draw (0,0) node {$\blacktriangleleft$};
\draw (4,0) node {$\blacktriangleright$};

\draw (2, .75) node {\scalebox{1.5}{$\bullet$}};
\draw (2,-.75) node {\scalebox{1.5}{$\bullet$}};

\end{tikzpicture}
}
\ee
Our primary focus will be on non-stochastic counterterms since in our universe the stochastic ones are expected to be sub-dominant at long wavelengths (see below).

%%%%%%%%%%%%%%%%%%%%%%%%%%%%%%%%%%%%%%%%%%%%%%%%%%%%%
\subsection{Example: One-loop power spectrum}

As an explicit example let us consider the 1-loop non-stochastic contribution to the power spectrum \eqref{P13}. This corresponds to 
\be\label{P1PI}
\raisebox{-21 pt}{
\begin{tikzpicture}[scale=.7]

\draw[very  thick] (-1,0) -- (1,0);
\draw [very  thick](3,0) -- (5,0);
\draw [very  thick](3,0) arc (0:-180:1cm and .75cm);
\draw [very  thick](3,0) arc (0:180:1cm and .75cm);

\draw[-{triangle 45}] (2,.75) arc (90:35:1cm and .75cm) ;
\draw[-{triangle 45}] (2,.75) arc (90:145:1cm and .75cm) ;

\draw (2,-.75) node {$\blacktriangleleft$};
\draw (0,0) node {$\blacktriangleleft$};
\draw (4,0) node {$\blacktriangleleft$};

\draw (2,.75) node {\scalebox{1.5}{$\bullet$}};
\draw (.5,-.6) node {$\eta_1,a_1$};
\draw (3.7,-.6) node {$\eta_2,a_2$};
%\draw (-1.6,0) node {$\eta,a$};
\draw (5.7,0) node {$\eta',b$};

\end{tikzpicture}
}
\ee
whose amplitude is given by (recall that amputation removes the final propagator)
\be\label{amp2pf}
\int^{\eta_1}_{\eta'} d\eta_2~ C_{a_1}(\eta_1;\k,a_2,\eta_2)
~g_{a_2\,b}(\eta_2,\eta')\,.
\ee
Here
\ba
C_{a}(\eta_1;\k,b,\eta_2) & = &
4g_{a_2\,1}(\eta_1,\eta_0) g_{b_2\,1}(\eta_2,\eta_0) g_{a_1\,b_1}(\eta_1,\eta_2)  \\
& & \times \int_\q P_{\rm lin}(q)\:\gamma _{a\,a_1\,a_2}\left(\K,+\Q,\K-\Q\right)\: \gamma_{b_1\,b_2\,b}\left(\K-\Q,-\Q,\K\right) \,, \nonumber
\ea
the time variable is defined as $a=e^\eta$, $g_{a_2 1}(\eta_1,\eta_0)=g_{a_2 2}(\eta_1,\eta_0)$ is the same as the growth factor, and the summation of repeated indices is assumed. Using the explicit expressions for the propagator and vertices, and restricting to $q\gg k$, we obtain the following expression for the ultraviolet part of the loop diagram
\ba
\label{CUV}
C^{UV}_{a}(\eta_1;\k,b,\eta_2)= ~ k^2 \sigma^2_v ~~ e^{-\eta_1/2}\Bigg\{
\Bigg[  \begin{array}{ccc}
\frac{3}{5} & \frac{-4}{25}
\\
\frac{-9}{5} & \frac{-8}{5}
\end{array}
\Bigg]~  e^{\frac{5 \eta_1 }{2} }+
%%%
\Bigg[  \begin{array}{ccc}
\frac{-8}{5} & \frac{4}{25}
\\
\frac{9}{5} & \frac{3}{5}
\end{array}
\Bigg] ~e^{\frac{5 \eta_2 }{2}} \Bigg\}\,.
\ea 
The velocity dispersion, which is defined as
\be
\sigma_v^2 \equiv \int^\Lambda dq P_{\rm lin}(q),
\ee
is the cutoff dependent coefficient that we aim to renormalize by the addition of a counterterm proportional to $k^2$. This will be done in the next section.

%%%%%%%%%%%%%%%%%%%%%%%%%%%%%%%%%%%%%%%%%%%%%%%
\section{Renormalization and counterterms}\label{counter}

In this section, we first explain how to read off the counterterm associated to each amputated loop graph. Then, we discuss various properties of these counterterms, give their general structure, and conclude with an explicit application of these ideas to the 1-loop bispectrum.

The amplitude of a generic amputated diagram (such as \eqref{t1PIa}) with the final vertex inserted at time $\eta$, outgoing momentum and index $(\k,a)$, and a set of soft ingoing lines characterized by $\{\k_n,b_n,\eta'_n\}$, and hard ingoing lines characterized by $\{\q_i\}$ (recall that hard lines are all initial fields $\delta_1(q_i)$, so $b_i$ and $\eta'_i$ need not be specified), is of the form
\be
\int^{\eta}_{\eta_0}\{d\eta_n\} A_{a a_1\cdots a_n}[\eta;\{\k_n,\eta_n\};\{\q_i\}]
\prod_n g_{a_n b_n}(\eta_n,\eta'_n)\prod_i \delta_1(\q_i).
\ee
In the non-stochastic case the hard lines $\{\q_i\}$ can be paired and contracted to give $P_{\rm lin}(q_i)$ and integration over their momenta results in
\be
\int^\eta_{\eta_0}\{d\eta_n\} C_{a a_1\cdots a_n}[\eta;\{\k_n,\eta_n\}]
\prod_n g_{a_n b_n}(\eta_n,\eta'_n).
\ee
The $C_{ab}$ of last section was an example of this with only one ingoing soft line. The $C$ kernel is a cutoff dependent quantity. This diagram is renormalized by the addition of a counterterm of the following form to the equation of motion (c.f. \eqref{t1PIb})
\be\label{ct}
\int^\eta_{\eta_0}\{d\eta_n\} c_{{a a_1\cdots a_n}}[\eta;\{\k_n,\eta_n\}]\prod_n \Psi_{a_n}(\k_n,\eta_n).
\ee
To cancel the cutoff dependence of the loop diagrams, the dependence of $c_{a\dots}$ on $\{\k_n\}$ has to match that of $C^{UV}_{a\cdots}$, namely the cutoff dependent part of $C_{a\cdots}$ when it is expanded in powers of $k/q$. Let us analyze this more closely. In the ultraviolet regime the momentum dependence of $C_{a\cdots}$ can always be brought via Taylor expansion into a sum of terms, each of which factorizes into a product of a function of $\{\q_i\}$ times a function of $\{\k_n\}$. We are interested in the functional dependence on $\{\k_n\}$. Consider a vertex at $\eta_n$ where the $n^{th}$ soft line is attached to the diagram. From the equations of motion \eqref{sptcoupled} and \eqref{thetaeq}, the leading order vertices of the theory are 
\be\label{vertices}
-v^i\d_i \delta,\quad -\theta \delta,\quad -v^i\d_i \theta,\quad -\left( \d_iv^j \right) \d_j v^i.
\ee
Thus, the soft line can realize one of the following six possibilities:
\be
\begin{split}
v^i\to \frac{k_n^i}{k_n^2}\Psi_2(\k_n,\eta_n), \quad \d_i\delta\to k_n^i\Psi_1(\k_n,\eta_n),
\quad \theta\to \Psi_2(\k_n,\eta_n),\\[10pt]
\delta\to \Psi_1(\k_n,\eta_n),\quad \d_i\theta\to k_n^i\Psi_2(\k_n,\eta_n),
\quad \d_i v^j\to \frac{k_n^ik_n^j}{k_n^2}\Psi_2(\k_n,\eta_n),
\end{split}
\ee
These are the only ways in which $c_{a\cdots}$ may depend on the $n^{th}$ field, up to possible multiplication by positive factors of $k_n^i$, arising from vertices appearing at later times in the diagram. The contribution from these later vertices is guaranteed to be analytic in $k_n^i$ because (i) by definition of an amputated graph, a soft line $n$ attaches immediately to a hard line, say with momentum $\q_n$, thus (ii) the later vertices depend on $\k_n$ only through $\q_n+\k_n$, and (iii) since $k_n\ll q_n$ one can Taylor expand in $k_n^i$. 

Moreover, $c_{a\cdots}$ has to have a cutoff dependent coefficient with exactly the same time-dependence and overall amplitude as the one arising from the integration over $\{\q_i\}$ in $C^{UV}_{a\cdots}$. However, there can be a cutoff independent piece with different time-dependence. This counterterm by construction renormalizes all possible embeddings of the original amputated diagram into any higher order diagram. Next, we will discuss the manifestation of characteristic properties of the cosmic fluid in the form of these new, and yet rather abstract, corrections.
 
%%%%%%%%%%%%%%%%%%%%%%%%%%%%%%%%%%%%%%%%%%%%%%%%%%%%%%%
\subsection{Long memory effect and non-locality in time}\label{memory}

Counterterms capture the effect of the short wavelength modes on the long wavelength physics. They result from replacing hard loop diagrams with new effective interactions among soft lines (i.e.~integrating out the short-distance physics). The result of this procedure is guaranteed to be spatially local since the short modes can probe the long modes only in a region of size $\Lambda^{-1}$ over which the long modes can be accurately Taylor expanded in powers of $k/\Lambda$. However, before reaching the virialization scale, the short modes evolve with approximately the same time scale $\cH^{-1}$ as the long modes. Hence, we should expect the counterterms to be generically non-local in time as a short mode can be influenced by long wavelength modes at some earlier time and respond at a much later time. The hard lines in a loop diagram can therefore be collapsed to a point spatially but not temporally \cite{Carroll,2loop}. 

This expectation is manifest in our general formula \eqref{ct} and in the explicit 1-loop example. In order to systematically renormalize the amputated diagram of \eqref{amp2pf} one needs to add a temporally nonlocal quadratic vertex, corresponding to a linear counterterm in the equations of motion:
\be\label{nlt}
\partial_{\eta} \Psi_a(\k,\eta) + \Omega_{a\,b} \Psi_b(\k,\eta) 
= k^2 \int^{\eta}_{0} d\eta_1 ~c_{ab}(\eta;\eta_1)~~\Psi_{b}(\k,\eta_1)+\cdots\,,
\ee
where we omitted the standard cubic vertices. Here $k^2 c_{ab}$ has a cutoff dependent piece that (up to a minus sign) exactly matches $C^{UV}_{ab}$ in \eqref{CUV} so that the sum of the two cancels, up to a regularization independent piece. As will be seen in section \ref{sec:1PR} this counter-term automatically renormalizes the 1PR contribution \eqref{B1PR} to the bispectrum. On the other hand, if we were only interested in renormalizing the $P_{13}$ diagram \eqref{P13}, this vertex could be replaced by a local one (the conventional speed of sound and viscosity counterterms)
\be\label{near}
\tilde c_a(\eta) k^2 \Psi_1(\eta),
\ee
where
\be
\tilde c_a(\eta)\equiv \frac{1}{D^{+}(\eta)}\sum_b \int^{\eta}_{0} d\eta_1 ~c_{aa_1}(\eta;\eta_1)~~D^{+}_{a_1 b}(\eta_1).
\ee
As we will see in section \ref{sec:remedy} this reduction is always possible in perturbation theory. However, it is incompatible with systematic renormalization. For example, using the local-in-time counterterm in  \eqref{near}, on finds that the 1PR diagram \eqref{B1PR} requires new counterterms \cite{Baldauf}. 

%%%%%%%%%%%%%%%%%%%%%%%%%%%%%%%%%%%%%%%%%%%%%%%%%%%%%%
\subsection{Large scale flow and the Equivalence Principle\label{equiv}}

The origianl system \eqref{sptcoupled} is invariant under the addition of a uniform bulk flow \cite{Kehagias,Peloso}
\be\label{EP}
\x \to \x + \n(\tau),\qquad \v \to \v + \dot\n(\tau),
\ee
where $ \n $ is an arbitrary time-dependent vector. This ensures that a very long-wavelength perturbation, whose main effect on much shorter scales is to produce a uniform acceleration, is locally unobservable, in accord with the Equivalence Principle.\footnote{For a discussion of these symmetries from the perspective of Weinberg adiabatic modes see appendix A of \cite{Baldauf}.} Local observables can depend on the tidal field $\d_i\d_j\phi$, the shear $\d_i v_j$, their spatial derivatives, and their convective time-derivatives
\be\label{convective}
D_\tau \equiv  \d_\tau +\v\cdot \nabla.
\ee
The counterterms which characterize the response of the short scale physics to long-wavelength modes should also follow the same rules. The short modes can only respond to the locally measurable quantities. However, when the counterterms are non-local in time as in \eqref{nlt}, they depend on these local observables along the large scale flow. The same considerations imply that the time integral must be taken along the fluid trajectory, which is implicitly defined by 
\be
\xfl[\tau';\x,\tau] = \x -\int^\tau_{\tau'} d\tau'' \v(\xfl[\tau'';\x,\tau],\tau'').
\ee
In the following, we often drop the last two arguments of $\xfl$ when no ambiguity arises. For concreteness, let us focus on the counterterm in \eqref{nlt}. To be invariant under \eqref{EP}, it must be dressed to read \cite{2loop}
\be\label{ctxfl}
\nabla^2\int^{\eta}_{\eta_0} d\eta' ~c_{ab}(\eta;\eta')~~\Psi_{b}(\xfl(\eta'),\eta').
\ee
The significance of this replacement $\x\to \xfl$ soon becomes clear in the example of 1-loop bispectrum in section \ref{1PI-shift-sec}. Perturbatively expanding $\xfl$ around $\x$, we get
\be\label{shift}
\begin{split}
\Psi_a(\xfl[\eta';\x,\eta],\eta')=&\Psi_a(\eta')-\int^{\eta}_{\eta'} d\eta_1 \v(\eta_1)\cdot\nabla \Psi_a(\eta')\\[10pt]
&+\int^{\eta}_{\eta'}d\eta_1 \int^{\eta}_{\eta_1} d\eta_2 \ [\v(\eta_2)\cdot\nabla\v(\eta_1)]\cdot\nabla \Psi_a(\eta')+\cdots
\end{split}
\ee
where all omitted space arguments are now $\x$. This replacement essentially connects an infinite set of counterterms (and hence an infinite set of amputated diagrams) to one another. These are all diagrams in which the original diagram \eqref{Psi3} is being displaced by the bulk motion produced by any number of ingoing soft lines attached at later times than the original attachment of $\k_1$:
\be\label{vPsi3}
\raisebox{-65 pt}{
\begin{tikzpicture}[scale=.7]
\draw [line width= 2 pt](0,0) -- (4.037,4.037);
\draw [line width= 2 pt](4,4) -- (8,0);

\draw [line width= .75 pt] (4,4) -- (4,5.5);
\draw [line width= .75 pt] (1,1) -- (2,0);
\draw [line width= .75 pt] (2,2) -- (2.5,1.5);
\draw [line width= .75 pt] (3,3) -- (3.5,2.5);
\draw [line width= .75 pt] (6.25,0.75) -- (6.75,1.25);
\draw [line width= .75 pt] (5,2) -- (5.5,2.5);

\draw (2.75,1.25) node  {$v$};
\draw (3.75,2.25) node  {$v$};
\draw (6.,.5) node  {$v$};
\draw (4.75,1.75) node  {$v$};

\draw (0,-.5) node  {$\Q$};
\draw (2,-.5) node  {$\K_1$};
\draw (8,-.5) node  {$-\Q$};
\draw (3.5,5) node  {$\K$};

\end{tikzpicture}
}
\ee
More generally, whenever a soft line realizes the $v^i\d_i$ part of the vertices in \eqref{vertices}, it can be attributed to the shift from $\x$ to $\xfl$ of a lower order diagram with the soft line removed. These diagrams have the following characteristic property in perturbation theory: In the presence of a long-wavelength mode with constant amplitude $\delta_1(\k_L)$ and vanishing $k_L$, they diverge as $\k_L/k_L^2$, namely, proportionally to the bulk velocity induced by the long mode. This will be referred to as infra-red (IR) singularity. The UV part of these diagrams are fully fixed in terms of that of the original (unshifted) diagram. The counter-terms associated to these diagrams too are fully fixed in terms of that of the original diagram. Therefore, they do not introduce any new parameters into the effective theory. Eliminating these diagrams by shifting $\x\to\xfl$ ensures that the new higher order counterterms needed to renormalize the remaining higher order 1PI diagrams depend only on locally observable quantities, again measured along the dark matter flow $\xfl$. In other words there is no IR singularity, except in the argument $\xfl$ of the fields.\footnote{This fixed IR singular part can be used to derive a universal and dominant piece of the squeezed correlation functions, when one of the external momenta is much smaller than the others \cite{Creminelli}. This piece often vanishes in equal-time correlation functions, however there is an exception in the presence of BAO feature \cite{BAO}.}

%%%%%%%%%%%%%%%%%%%%%%%%%%%%%%%%%%%%%%%%%%%%%%%%%%%%%%%
\subsection{Soft outgoing momentum and vorticity}\label{sec:soft}

Another property of loop diagrams to be discussed is their softness in the outgoing momentum. In perturbation theory, it can be easily verified that all vertices are at least single soft:
\be\label{k2}
\lim_{q\to \infty} \gamma_{abc}(\k,\q,\k-\q) = \O(k/q).
\ee
In any amputated loop diagram, the latest vertex is one in which two hard lines combine to give an outgoing soft line of momentum $k$ (see e.g. \eqref{t1PIa}). Because of \eqref{k2} the UV part of this diagram will contain at least one factor of $k$, and so do the corresponding counterterms. Rotational invariance implies that the full amputated diagram must be of second order in soft momenta since there is no singularity (i.e.~negative power) in the ingoing soft momenta after taking care of the shift terms as explained in the previous section. Thus expressed in terms of $\Psi_a$, the counterterms must start from second order in derivatives, one of them being an overall derivative because of the property \eqref{k2} of the final vertex. An example is
\be\label{cubic}
\d_i[\Psi_{a_1}(\xfl(\eta_1),\eta_1)\Psi_{a_2}(\xfl(\eta_2),\eta_2)\d_i\Psi_{a_3}(\xfl(\eta_3),\eta_3)].
\ee
The ultraviolet part of $C_{ab}(\eta;\k,\eta)$, calculated above agrees with this expectation.

The above softness property of the counter-terms has a natural explanation based on the structure of fluid equations. The continuity equation relates the change in mass density to the divergence of the mass current (momentum density)\footnote{Notice that we have changed the traditional definition of $ \pib $ by a factor of $ 1/\bar \rho $, in such a way that $ \pib $ is dimensionless. This is convenient because $ \bar \rho $ completely disappears from all the equations.} 
\be\label{cont}
\dot \delta =-\nabla\cdot \pib,\qquad \text{where} \quad \pib \equiv (1+\delta)\v.
\ee
However, this relation between $\pib$ and $\v$ depends on the definition of velocity. Fixing velocity at some scale $\Lambda$ to be given by $[\pi]_\Lambda/[1+\delta]_\Lambda$ and then coarse-graining to a smaller cutoff $\Lambda'$ results in a velocity field that differs from $[\pi]_{\Lambda'}/[1+\delta]_{\Lambda'}$. Hence, when working with velocity field as the primary variable one must add a locally observable vector $\bsb j_c$ as a counterterm to the mass current. The absence of IR singularity ensures that a locally observable vector must be at least of first order in derivatives (in the same sense that \eqref{cubic} is of second order in derivatives). Upon taking the divergence, this leads to a counterterm $\nabla \cdot \bsb j_c$ in the continuity equation that is second order in derivative counting, and because of the overall derivative it contains at least one factor of the outgoing momentum. 

Similarly, integrating out the short wavelength modes can introduce an effective force $\bsb f_c$ in the Euler equation. This is again a locally observable vector that incorporates the counterterms. After taking the divergence to obtain an equation for $\theta$ it leads to a counterterm $\nabla \cdot (\bsb f_c/(1+\delta))$ with an overall derivative. 

It is worth noting that since the original system of equations \eqref{sptcoupled} do not generate any vorticity, to cancel the cutoff dependence of just SPT loops, $\bsb f_c/(1+\delta)$ can be taken to be curl-free. In other words, the counterterms to the $\theta$ equation $c_{a=2,\cdots}$ are not only a total divergence but they are the Laplacian of locally observable scalars. We will verify this expectation in the explicit calculation of quadratic counterterms (see appendix \ref{app:2ndct}). Of course, the finite pieces of the counterterms do not have to obey this rule, and in principle the short scale dynamics can generate vorticity (see section \ref{velind} and appendix \ref{app:v} for comments on the non-uniqueness of the definition of velocity). 

The property \eqref{k2} and the fact that all counterterms have one overall derivative ensures that if the outgoing momentum $\k$ is much smaller than the soft ingoing momenta $\{k_i\}$ the solution for $\delta(\k)$ and $\theta(\k)$ is going to be single-soft in $k$. In fact, as we will discuss in section \ref{double}, momentum conservation implies that $\delta$ must not only be single-soft but rather double-soft in $k$ in the limit $k\ll k_i$. As a result, only a subset of locally observable forces -- those that respect this property -- are allowed to be added as counterterms to the system. 

%%%%%%%%%%%%%%%%%%%%%%%%%%%%%%%%%%%%%%%%%%%%%%%%%%%%
\subsection{The list of counterterms}\label{list}

Using the above rules we can write down an over-complete list of counterterms which can be added to the equations of motion. The softness in the outgoing line implies that they must be of the form $\d_i c^{i}$. $c^{i}$ encodes the memory of all incoming soft lines that are attached to the hard lines of the amputated diagrams \eqref{t1PIa}. Therefore, each field in $c^{i}(\x,\eta)$ is integrated from $\eta_0$ to $\eta$ against a general time-dependent memory function. Any attached ingoing line is either a locally measurable quantity made of the soft mode such as $\d_i\d_j\phi$ or $\d_iv_j$ (and their convective time derivatives \eqref{convective}), or it is a velocity, which is not a local observable. As argued above the latter is necessary to perturbatively dress the time integrals at constant $\x$ to those along the fluid flow. In the former case, since both $\d_i\d_j\phi$ and $\d_i v_j$ have two indices, rotational invariance implies that to construct a vector $c^i$ one needs to have one additional derivative.

We proceed as follows. For simplicity, suppose there is zero vorticity; this allows us to write $\v =\nabla \phi_v$, and introduce 
\be
\phi_a \equiv \nabla^{-2}\Psi_a = (\phi,\phi_v).
\ee
Next, as an intermediate step we construct $n^{th}$ order two-index tensors $c^{ij}$ made of various contractions of the local observables $\Pi_a^{ij}\equiv \d_i\d_j\phi_a$. Using matrix notation, the first few orders are
\ba
&{\rm 1^{st}:}& \Pi_a,\unit \tr{\Pi_a}\\ \label{2nd}
&{\rm 2^{nd}:}& \Pi_a \tr(\Pi_b), \Pi_a\Pi_b, \unit\tr(\Pi_a)\tr(\Pi_b),\unit\tr(\Pi_a\Pi_b) \\
&{\rm 3^{rd}:}& \Pi_a \tr(\Pi_b)\tr(\Pi_b),\Pi_a\tr(\Pi_a\Pi_b),\Pi_a\Pi_b \tr(\Pi_c),\Pi_a\Pi_b\Pi_c,\cdots
\ea
where $\unit^{ij}=\delta^{ij}$, and as an example, $\tr(\Pi_a\Pi_b)=\sum_{ij}\Pi_a^{ij}\Pi_b^{ji}$. Each $\Pi_a$ is evaluated at a different time which is integrated against a temporal kernel until the final time. An over-complete (see below) list of vectors $c^i$ consists of all possible ways of acting by $\d_j$ on any of $\Pi_a$ appearing in $c^{ij}$. Finally, one takes the divergence $\d_ic^i$. One second-order example is
\be\label{c2}
\d_i\int^\eta_{\eta_0} d\eta_1 d\eta_2 K_{a a_1 a_2}(\eta,\eta_1,\eta_2)
\d_j\Pi_{a_1}^{ik}(\xfl(\eta_1),\eta_1) \Pi_{a_2}^{jk}(\xfl(\eta_2),\eta_2),
\ee
where the derivative in $\d_j\Pi_{a_1}$ should, in principle, be with respect to $\xfl$. However, the difference between this and the derivative with respect to $\x$ can be expressed in terms of higher order terms that are included in our list -- in other words 
\be
\frac{\d x_{\rm fl}^i}{\d x^j} = \delta^{ij} +\text{local observable tensors.}
\ee
So we take it to be $\d/\d x^j$ in what follows. Note also that the arbitrary time-kernel makes the use of convective time-derivatives redundant. 

Unfortunately the above procedure generates an \textit{over-complete list of counter-terms}, as it allows us to introduce arbitrary observable vectors $\bsb f_c$ and $\bsb j_c$ as counter-terms. On the other hand, as we will argue in section \ref{double}, only a subset of these vectors -- those which ensure double softness of $\delta$ when $k_{\rm out}\ll \{k_i\}_{\rm in}$ -- are permitted by momentum conservation. An allowed list can be obtained by restricting the kernels to those which arise from perturbation theory. This elects certain combination of these vectors that are necessary to cancel the cutoff dependence of the loops. These combinations inherit the double softness property of the leading order perturbation theory which is discussed in section \ref{double_pert}. However, in this way all amputated diagrams with fixed number of ingoing soft and hard lines combine to give only one new counter-term whose coefficient can be allowed to have a cutoff independent piece. While this is sufficient to cancel the cutoff dependence of the loops, in order to produce a complete basis that can give an effective description for all possible microscopic scenarios one needs to proceed to diagrams with increasingly large number of hard modes. 

We will next verify the adequacy of this formalism in cancelling the cutoff dependence in the explicit example of the 1-loop bispectrum. Several properties of the counterterms become apparent in this example. In section \ref{sec:remedy}, we show how this rather cumbersome and nonlocal formulation can be reduced to a much more tractable and local one in any practical calculation by giving up systematic renormalization. Moreover, using the results of section \ref{double}, we will provide an alternative list of local counterterms that is complete but not over-complete.

%%%%%%%%%%%%%%%%%%%%%%%%%%%%%%%%%%%%%%%%%%%%%%%%%%%%%%%%%%%
\subsection{Example: One-loop bispectrum}

In this section we discuss the renormalization of the non-stochastic part of the 1-loop bispectrum which is conventionally called $B_{114}$. The 1PI and 1PR contributions are evaluated and discussed separately.

%%%%%%%%%%%%%%%%%%%%%%%%%%%%%%%%%
\subsubsection{1PR diagrams}\label{sec:1PR}

As expected from the general argument at the beginning of this section, it is easily seen that the 1PR diagrams are automatically renormalized by the insertion of the linear counterterm in \eqref{nlt} in a tree-level diagram. The diagram
\be\label{B1PR1}
\raisebox{-56 pt}{
\begin{tikzpicture}[scale=.7]
\draw [very  thick](-.5,0) -- (1,0);
\draw [very  thick](3,0) -- (4,0);
\draw [very  thick](3,0) arc (0:-180:1cm and .75cm);
\draw [very  thick](3,0) arc (0:83:1cm and .75cm);
\draw [very  thick](1,0) arc (180:97:1cm and .75cm);
\draw [very  thick](4,0) -- (6,2);
\draw [very  thick](4,0) -- (6,-2);

\draw[-{triangle 45}] (2,.75) arc (90:35:1cm and .75cm) ;
\draw[-{triangle 45}] (2,.75) arc (90:145:1cm and .75cm) ;
\draw (3.5,0) node {$\blacktriangleleft$};
\draw (2,-.75) node {$\blacktriangleleft$};
\draw (.25,0) node {$\blacktriangleleft$};
\draw (4.5,.5) node [rotate=45] {$\blacktriangleleft$};
\draw (4.5,-.5) node [rotate=-45] {$\blacktriangleleft$};
\draw (5.5,1.5) node [rotate=45] {$\blacktriangleright$};
\draw (5.5,-1.5) node [rotate=-45] {$\blacktriangleright$};

\draw (5,1) node {\scalebox{1.5}{$\bullet$}};
\draw (5,-1) node {\scalebox{1.5}{$\bullet$}};
\draw (2,.75) node {\scalebox{1.5}{$\bullet$}};

\draw (-1.6,0) node {$\k,a;\eta$};
\draw (.8,-.3) node {\scalebox{.7}{$a_1$}};
%\draw (.8,0) node {\scalebox{.8}{$\square$}};
\draw (3.3,-.3) node {\scalebox{.7}{$a_2$}};
\draw (1.3,0) node {\scalebox{.7}{$\eta_1$}};
\draw (2.7,0) node {\scalebox{.7}{$\eta_2$}};
\draw (6.5,2.5) node {$\k_1,b;\eta$};
\draw (6.5,-2.5) node {$\k_2,c;\eta$};

\end{tikzpicture}}
\ee
corresponds to taking the expectation value
\be\label{1pr1}
\int^\eta_{\eta_0} d\eta_1 g_{a a_1}(\eta,\eta_1)\int^{\eta_1}_{\eta_0} d\eta_2~ C_{a_1 a_2}(\eta_1,\eta_2;\k) 
\expect{\Psi_{a_2}^{(2)}(\k,\eta_2)\Psi_b^{(1)}(\k_1,\eta)\Psi_c^{(1)}(\k_2,\eta)}+\text{2 perm.}
\ee
We denoted the UV part of this loop diagram by $C^{UV}$ and it will be renormalized by adding the contribution of the diagram 
\be\label{ctB1PR1}
\raisebox{-56 pt}{
\begin{tikzpicture}[scale=.7]
\draw [very  thick](-1,0) -- (.3,0);
\draw [very  thick](0.7,0) -- (2,0);
\draw [very  thick](2,0) -- (4,2);
\draw [very  thick](2,0) -- (4,-2);

\draw (-.25,0) node {$\blacktriangleleft$};
\draw (1.25,0) node {$\blacktriangleleft$};
\draw (2.5,.5) node [rotate=45] {$\blacktriangleleft$};
\draw (3.5,1.5) node [rotate=45] {$\blacktriangleright$};
\draw (2.5,-.5) node [rotate=-45] {$\blacktriangleleft$};
\draw (3.5,-1.5) node [rotate=-45] {$\blacktriangleright$};

\draw (3,1) node {\scalebox{1.5}{$\bullet$}};
\draw (3,-1) node {\scalebox{1.5}{$\bullet$}};
\draw (.5,0) node {$\bm{\otimes}$};

\draw (-2,0) node {$\k,a;\eta$};
%\draw (-.3,-.5) node {\scalebox{.7}{$\eta_1,a_1$}};
%\draw (1.4,-.5) node {\scalebox{.7}{$\eta_2,a_2$}};
\draw (4.5,2.5) node {$\k_1,b;\eta$};
\draw (4.5,-2.5) node {$\k_2,c;\eta$};

\end{tikzpicture}}
\ee
where the cross represents the counterterm in \eqref{nlt}. The expression for this diagram is given by an identical expression as \eqref{1pr1} with $C\to k^2 c$, which have equal and opposite cutoff dependences. The other 1PR diagram 
\be\label{B1PR2}
\raisebox{-56pt}{
\begin{tikzpicture}[scale=.7]
\draw [very  thick](0,0) -- (2,0);
\draw [very  thick](2,0) -- (4,-2);
\draw [very  thick](2,0) -- (3,1);
\draw [very  thick](3,1)[rotate=45] arc (-180:0:1cm and .75cm);
\draw [very  thick](3,1)[rotate=45] arc (180:90:1cm and .75cm) ;
\draw [very  thick](4.42,2.42)[rotate=45] arc (0:78:1cm and .75cm);
\draw [very  thick](4.42,2.42) -- (6.2,4.2);
%\draw [blue](1,0) arc (180:90:1cm and 1cm);
\draw (3.25,2.25) node {\scalebox{1.5}{$\bullet$}};
\draw (3,-1) node {\scalebox{1.5}{$\bullet$}};
\draw (5.3,3.3) node {\scalebox{1.5}{$\bullet$}};

\draw[-{triangle 45}] (3.18,2.25) [rotate=45] arc (90:35:1cm and .75cm) ;
\draw[-{triangle 45}] (3.18,2.25)[rotate=45] arc (90:145:1cm and .75cm) ;

\draw (1,0) node {$\blacktriangleleft$};
\draw (2.5,.5) node [rotate=45] {$\blacktriangleleft$};
\draw (4.2,1.16) node [rotate=45] {$\blacktriangleleft$};
\draw (2.5,-.5) node [rotate=-45] {$\blacktriangleleft$};

\draw (5.9,3.9) node [rotate=45] {$\blacktriangleright$};
\draw (4.8,2.8) node [rotate=45] {$\blacktriangleleft$};

\draw (3.5,-1.5) node [rotate=-45] {$\blacktriangleright$};

\draw (-1,0) node {$\k,a;\eta$};
\draw (2.7,0) node {\scalebox{.7}{$\eta_1$}};
\draw (3.2,1.25) node [rotate=45]{\scalebox{.7}{$\eta_2$}};
\draw (4.2,2.2) node [rotate=45]{\scalebox{.7}{$\eta_3$}};
\draw (2.1,.5) node [rotate=45]{\scalebox{.7}{\textcolor{black}{$b_1$}}};
\draw (2.6,1.05) node [rotate=45]{\scalebox{.7}{\textcolor{black}{$b_2$}}};
\draw (4.4,2.85) node [rotate=45]{\scalebox{.7}{\textcolor{black}{$b_3$}}};
\draw (2.1,-0.41) node [rotate=-45]{\scalebox{.7}{\textcolor{black}{$c_1$}}};
\draw (1.7,.2) node {\scalebox{.7}{\textcolor{black}{$a_1$}}};
\draw (6.8,4.6) node {$\k_1,b;\eta$};
\draw (4.5,-2.5) node {$\k_2,c;\eta$};

\end{tikzpicture}}
\ee
has an expression
\be\label{1pr2}
\begin{split}
\int\!\! d^3\q &\int^\eta_{\eta_0}\shnk d\eta_1\int^{\eta_2}_{\eta_0}\shnk d\eta_2\int^{\eta_2}_{\eta_0}\shnk d\eta_3\:
g_{aa_1}(\eta,\eta_1)\,g_{b_1b_2}(\eta_1,\eta_2) \:\gamma_{a_1\,b_1\, c_1 }(\k,\k_1,\k_2)\: C_{b_2 b_3}(\eta_2,\eta_3;\k_1) \\[10pt]
& \expect{\Psi_b^{(1)}(\k_1,\eta)\Psi_{b_3}^{(1)}(\k_1,\eta_3)\Psi_c^{(1)}(\k_2,\eta)\Psi_{c_1}^{(1)}(\k_2,\eta_1)}+\text{5 perm.}
\end{split}
\ee
Similarly, the UV part of this diagram $C^{UV}$ is renormalized by the diagram containing the same counterterm in place of the loop:
\be\label{ctB1PR2}
\raisebox{-66pt}{
\begin{tikzpicture}[scale=.7]
\draw [very  thick](0,0) -- (2,0);
\draw [very  thick](2,0) -- (2.84,.84);
\draw [very  thick](3.16,1.16) -- (4.75,2.75);
\draw [very  thick](2,0) -- (4.5,-2.5);

\draw (1,0) node {$\blacktriangleleft$};
\draw (2.5,.5) node [rotate=45] {$\blacktriangleleft$};
\draw (3.5,1.5) node [rotate=45] {$\blacktriangleleft$};
\draw (4.5,2.5) node [rotate=45] {$\blacktriangleright$};
\draw (2.5,-.5) node [rotate=-45] {$\blacktriangleleft$};
\draw (4,-2) node [rotate=-45] {$\blacktriangleright$};

\draw (3,1) node[rotate=45] {$\bm{\otimes}$};
\draw (3.25,-1.25) node {\scalebox{1.5}{$\bullet$}};
\draw (4,2) node {\scalebox{1.5}{$\bullet$}};

\draw (-.9,0) node {$\k,a;\eta$};
\draw (5.35,3.2) node {$\k_1,b;\eta$};
\draw (5.25,-3) node {$\k_2,c;\eta$};
%\draw (2,1) node {\scalebox{.7}{$\eta_1,a_1$}};
%\draw (2.75,1.75) node {\scalebox{.7}{$\eta_2,a_2$}};

\end{tikzpicture}

}
\ee
which has the same expression as \eqref{1pr2} with $C\to k^2 c$.

It is worth emphasizing here that the temporal non-locality of counterterm $c_{a\,a'}(\eta;\eta',\k)$ is crucial for it to cancel the cutoff dependences of all 1PR loop integrals. This accommodates the possibility that the response of short modes should strongly depend on the time-dependence of the long modes. If only local in time counterterm are used, the UV divergences in 1PR diagrams are not generically canceled by the lower order counterterms, spoiling the systematic character of our renormalization procedure. This is expected and in agreement with the general argument presented in section \ref{memory} about the long memory effect.

%%%%%%%%%%%%%%%%%%%%%%%%%%%%%%%%%%%%%%%%%%%%%%%%%%%%%%%%%%%%%%%%%
\subsubsection{1PI diagrams: shift terms}
\label{1PI-shift-sec}

As discussed in section \ref{equiv}, some of the 1PI diagrams contributing to $B_{114}$, whose $\Psi_a^{(4)}(\k)$ field is of the form
\be\label{B1PIa}
\raisebox{-98 pt}{
\begin{tikzpicture}
\draw [very  thick](.1,0) -- (1,0);
\draw [very  thick] (1,0) -- (3,2);
\draw [very  thick](1,0) -- (3,-2);
\draw [very  thick](3,-2) -- (3,2);
\draw [very  thick](3,2) -- (4,3);
\draw [very  thick](3,-2) -- (4,-3);

\draw (.4,0) node {$\blacktriangleleft$};
\draw (3.4,2.4) node[rotate=45] {$\blacktriangleleft$};
\draw (3.4,-2.4) node[rotate=-45] {$\blacktriangleleft$};
\draw (2,-1) node[rotate=-45] {$\blacktriangleleft$};
\draw (2,1) node[rotate=45] {$\blacktriangleleft$};
\draw (3,1) node[rotate=90] {$\blacktriangleright$};
\draw (3,-1) node[rotate=90] {$\blacktriangleleft$};

\draw (3,0) node {\scalebox{1.5}{$\bullet$}};

\draw (-.5,0) node {$a, \K$};
\draw (1,-.5) node {$\eta$};
%\draw (.5,-.5) node{$\K$};
\draw (4.7,3.2) node {$a_1, \K_1$};
%\draw (3.8,2) node{$\K_1$};
\draw (2.8,2.5) node{$\eta_1$};
\draw (4.7,-3.2) node {$\mathbf{2},\K_2$};
%\draw (3.8,-2) node{$\K_2$};
\draw (2.8,-2.5) node{$\eta_2$};
\draw (7,.5) node{$ \K_1 \leftrightarrow \K_2$};
\draw (7,0) node{$\eta_1 \leftrightarrow \eta_2$};
\draw (7,-.5) node{$a_1 \leftrightarrow a_2$};
\draw (5.75,0) node{$+\Bigg ( $};
\draw (8,0) node{$\Bigg )$};

\end{tikzpicture}}
\ee
and 
\be\label{B1PIb}
\raisebox{-95 pt}{
\begin{tikzpicture}
\draw [very  thick](.1,0) -- (1,0);
\draw [very  thick](1,0) -- (3,2);
\draw [very  thick](1,0) -- (3,-2);
\draw [very  thick](3,2) -- (3,-2);
\draw [very  thick](3,2) -- (4,3);
\draw [very  thick](3,-2) -- (4,-3);

\draw (.4,0) node {$\blacktriangleleft$};
\draw (3.4,2.4) node[rotate=45] {$\blacktriangleleft$};
\draw (3.4,-2.4) node[rotate=-45] {$\blacktriangleleft$};
\draw (1.5,.5) node[rotate=45] {$\blacktriangleleft$};
\draw (2.5,1.5) node[rotate=45] {$\blacktriangleright$};
\draw (2,-1) node[rotate=-45] {$\blacktriangleleft$};
\draw (3,0) node[rotate=90] {$\blacktriangleleft$};

\draw (2,1) node {\scalebox{1.5}{$\bullet$}};

\draw (-.5,0) node {$a,\K$};
%\draw (.5,-.5) node{$\K$};
\draw (1,-.5) node {$\eta$};
\draw (4.7,3.2) node {$a_1,\K_1$};
%\draw (3.8,2) node{$\K_1$};
\draw (2.8,2.5) node{$\eta_1$};
\draw (4.7,-3.2) node {$\mathbf{2},\K_2$};
%\draw (3.8,-2) node{$\K_2$};
\draw (2.8,-2.5) node{$\eta_2$};
\draw (7,.5) node{$ \K_1 \leftrightarrow \K_2$};
\draw (7,0) node{$\eta_1 \leftrightarrow \eta_2$};
\draw (7,-.5) node{$a_1 \leftrightarrow a_2$};
\draw (5.75,0) node{$+\Bigg ( $};
\draw (8,0) node{$\Bigg )$};
\end{tikzpicture}
}
\ee
play the role of shifting the temporally nonlocal lower order counterterms to the fluid position (here $\mathbf{2}$ indicates the velocity field). It is implicit in the above diagrams that the lines marked by $\mathbf{2}$ realize $\v$ in $\v\cdot\nabla\Psi_a$ vertices, and that $\eta_2>\eta_1$. These correspond to contributions that diverge as $1/k_2$ when $k_2 \rightarrow 0$. We expect them to be renormalized by including the first shift term in \eqref{shift} in the counterterm for the 1-loop power spectrum \eqref{ctxfl}:
\ba\label{vpsi}
\nabla^2 \int^{\eta}_{\eta_0} d\eta_1 \int^{\eta}_{\eta_1} d\eta_2 ~c_{a\,b}(\eta,\eta_1) ~\dfrac{\nabla}{\nabla^2} 
\Psi_2(\x,\eta_2).\nabla\,\Psi_{b}(\x,\eta_1).
\ea
That this cancellation of IR-singularities must happen is almost obvious since any diagram that has a potential singularity (i.e.~any diagram where the soft line attached to the loop is a $v^i\d_i$) can be identified with a contribution to \eqref{vpsi}. An explicit check is provided in appendix \ref{app:IR}.

%%%%%%%%%%%%%%%%%%%%%%%%%%%%%%%%%%%%%%%%%%%%%%
\subsubsection{1PI diagrams: new counterterms}
\label{1PI-new-ct}
After the subtraction of shift terms from 1PI diagrams, what remains must be renormalized with the genuinely new higher order counterterms. These are made of locally observable operators integrated over the history of the short modes but evaluated at $\x$ (their displacement to $\xfl$ would correspond to higher order counterterms). At this level one must add a second order term to the equation of motion, corresponding to a cubic effective vertex. From \eqref{2nd}, there are 18 possibilities for quadratic terms in $c^{ij}$. Note that even if $a_1=a_2$ the two soft legs of the amputated diagram can be embedded in different ways in a connected diagram and be integrated against different time-kernels, therefore no symmetrization is needed. When taking derivatives to construct counterterms $\d_ic^i$, the number of operators proliferates but some are degenerate. It follows that, for fixed $a_{1,2}$ and $\eta_{1,2}$, possible $c^i$ vectors are
\be
\begin{split}
&\d_i\nabla^2\phi_{a_1}\nabla^2\phi_{a_2},\qquad \d_i\d_j\phi_{a_1}\nabla^2\d_j\phi_{a_2},
\qquad  \{a_1\leftrightarrow a_2,\eta_1\leftrightarrow \eta_2\},\\[10pt]
&\d_i\d_j\d_k\phi_{a_1} \d_j\d_k\phi_{a_2}+\{a_1\leftrightarrow a_2,\eta_1\leftrightarrow \eta_2\}
\end{split}
\ee
Thus there is a total of 15 independent, temporally nonlocal quadratic counterterms. It is shown in appendix \ref{app:2ndct} that they are sufficient to fully renormalize the 1PI diagrams. In fact, as expected from the structure of the leading vertices \eqref{vertices} and the requirement of momentum conservation (see the end of section \ref{list}), only a subset of these counterterms are allowed by the symmetries. The correct prescription will be discussed in section \ref{sec:remedy}. Also, because SPT does not generate vorticity (see subsection \ref{sec:soft}), an even smaller subset is needed to cancel UV-dependencies.

In summary, we have demonstrated that the effective theory of large scale structure can be systematically renormalized, at the expense of introducing a plethora of temporally nonlocal counterterms. After discussing the implications of locality and momentum conservation in the next section, in section \ref{sec:remedy} we explore a more practical approach.

%%%%%%%%%%%%%%%%%%%%%%%%%%%%%%%%%%%%%%%%%%%%%%%%%%
 
\section{Momentum conservation, locality and double softness}\label{double}

As alluded to in section \ref{sec:soft} momentum conservation and locality of the short scale dynamics ensure that the time evolution of short wavelength fluctuations can only lead to the generation of a longer wavelength perturbation suppressed at least by $k_{\rm out}^2$ \cite{Peebles}. In the following we will first review the heuristic argument of Peebles, and next show that double softness follows from the equations of motion of the fluid system. 

%%%%%%%%%%%%%%%%%%%%%%%%%%%%%%%%%%%%%%%%%%%%%%%%%%
 
\subsection{A general argument for double softness}

First consider an almost homogeneous initial density $\bar\rho_0$ with small short wavelength fluctuations of scale $1/q$, but no long wavelength fluctuation ($\delta(\k,\eta_0)=0$). Due to gravitational instability, these initial fluctuations collapse and form a clumpy distribution of matter $\rho(\x)$. We ask what is the typical size of a long wavelength fluctuation $\delta(\k,\eta)$ after this process? Approximating the final distribution by a set of point particles of mass $m_n$ at location $\x_n$, we have
\be\label{delta}
\delta(\k,\eta)=\frac{1}{\bar\rho}\sum_n m_n e^{i\k\cdot \x_n},
\ee
where $\bar\rho$ is the mean final density. If the short-scale dynamics was turned off, the local mass distribution would be uniform; we denote it by a tilde: $\tilde \rho(\x)=\bar\rho$. There is a nonzero $\delta(\k,\eta)$ because this would-be uniform distribution has clustered. in this hypothetical uniform universe, to keep track of the matter that ended up in each clump, we divide the space into (possibly overlapping) regions $R_n$ with density $\tilde\rho_n(\x)$. All of the mass in this region falls into the $n^{th}$ clump:
\be
m_n = \int_{R_n} d^3\y\  \tilde\rho_n(\x_n+\y).
\ee
If there is no overlap $\tilde\rho_n(\x)$ is the same as $\tilde\rho(\x)$ (and hence uniform) within $R_n$. But if two regions $R_{n_1}$ and $R_{n_2}$ overlap, then in the overlapping region $\tilde\rho_{n_1}(\x)+\tilde\rho_{n_2}(\x)=\bar\rho$ (and similarly for more overlaps). The typical size of each region is the extent to which mass elements are displaced from their initial position, namely, of order $1/q$. Now we can write
\be
\begin{split}
0&=\frac{1}{\bar\rho}\int d^3\x \bar\rho e^{i\k\cdot\x} 
= \frac{1}{\bar\rho}\sum_n \int_{R_n} d^3\y \tilde\rho_n(\x_n+\y) e^{i\k\cdot(\x_n+\y)} \\[10pt]
&=\frac{1}{\bar\rho}\sum_n e^{i\k\cdot\x_n}[m_n +ik^i d_n^i-k^i k^j Q_n^{ij}+\O(k^3/q^3)],
\end{split}
\ee
where $d_n^i$ and $Q_n^{ij}$ are, respectively, the first and second moments of the mass distribution in $R_n$ as measured from $\x_n$:
\be
\begin{split}
d_n^i &= \int_{R_n} d^3\y\ y^i\ \tilde\rho(\x_n+\y),\\[10pt]
Q_n^{ij}&=\int_{R_n} d^3\y\ y^i y^j\ \tilde\rho(\x_n+\y).
\end{split}
\ee
The expansion of the exponential is justified because of the smallness of the size of the region compared to $1/k$. Comparison with \eqref{delta} implies
\be\label{delta22}
\delta(\k,\eta) = -\frac{ik^i}{\bar\rho}\sum_n d_n^i e^{i\k\cdot\x_n}
+ \frac{k^ik^j}{\bar\rho}\sum_n Q_n^{ij}e^{i\k\cdot\x_n}+\O(k^3/q^3).
\ee
The second and higher order terms on the r.h.s. satisfy our general expectation about the dependence on $k$. As for the first term, the sum can be related to the center of mass position of each region:
\be\label{d}
\sum_n d_n^i e^{i\k\cdot\x_n}=\sum_n m_n(\x_n^{\rm CM}-\x_n)e^{i\k\cdot\x_n}.
\ee
The difference $(\x_n^{\rm CM}-\x_n)$ can only be caused by momentum transfer among different regions. If initially there existed only modes of wavelength $\sim 1/q$, there is no coherent momentum transfer over much longer distances, and the sum $\sum_n m_n(\x_n^{\rm CM}-\x_n)$ over any patch of size much larger than $1/q$ is negligible. Hence, the sum in \eqref{d} is nonzero to the extent that the exponential varies over any such patch, and therefore it must be suppressed at least by one factor of $k$. This shows that all terms in \eqref{delta22} are suppressed by at least two powers of $k$.\footnote{Our argument slightly differs from that of \cite{Peebles} in that the first term on the r.h.s. of \eqref{delta22} is absent there. However, this seems to be a nontrivial part of the argument.}

In the absence of long wavelength initial perturbations the location of the clumps $\x_n$ is effectively random at large scales, and the sums in \eqref{delta22} become identical to random-walks. Correlating two such contributions and averaging over the short modes results in 
\be
P_{\rm stochastic}(k\to 0)\propto k^4.
\ee
This is the generic asymptotic behavior of stochastic contributions to the long-wavelength power.

In the presence of initial long wavelength perturbations the above argument can be modified as follows. Now the clumpy universe should not be compared with a homogeneous universe, but with one in which the long modes are kept but the short modes are absent. Hence, $\tilde \rho(\x)$ is smooth but nonuniform. All other steps remain the same, though one should bear in mind that $\x_n$ and $\x_n^{\rm CM}$ both depend on and can get largely displaced by an initial long mode $\delta_1(\k_1)$; the leading dependence which goes as $\delta_1(\k_1)/k_1$ cancels in the difference in \eqref{d}. Nevertheless because of the remaining dependence of both $\x_n$ and the shape of each region $R_n$ on the initial long modes, \eqref{delta22} has nonzero correlation with the initial $\delta_1(\k_1)$. The non-stochastic diagram \eqref{P13} (equivalently \eqref{Psi3}) is one perturbative analog of such contribution.

Finally, a similar argument implies that the momentum density 
\be
\pib(\k,\eta) =\frac{1}{\bar \rho} \sum_n m_n \v_n e^{i\k\cdot \x_n}
\ee
starts from order $k$.

%%%%%%%%%%%%%%%%%%%%%%%%%%%%%%%%%%%%%%%%%%%%%%%%%%
\subsection{Double softness in perturbation theory}\label{double_pert}

The double softness property is not manifest in the formulation \eqref{sptcoupled} in terms of $\delta$ and $\v$. However, it becomes manifest if we eliminate $\v$ in favor of the momentum density $\pib=(1+\delta)\v$, in terms of which the continuity equation becomes linear \eqref{cont} and the Euler equation becomes
\be\label{momentum}
\dot\pi^i+\cH \pi^i = -(1+\delta)\d_i\phi -\d_j\left(\frac{\pi^i\pi^j}{1+\delta}\right).
\ee
Writing
\be
\delta\d_i\phi = \d_j(\d_j\phi \d_i\phi)-\frac{1}{2} \d_i(\d_j\phi \d_j\phi)
\ee
shows that the non-linear terms in this equation all have an overall derivative. Moreover, taking the divergence of this equation and using $\d_i\pi^i = -\dot \delta$ to obtain an equation for $\delta$ makes all interaction vertices explicitly second derivative. In appendix \ref{app:soft} it is shown that these equations evolve ingoing hard lines into a single soft answer for $\pib(\k_{\rm out})$ and a double soft answer for $\delta(\k_{\rm out})$ when $k_{\rm out}\ll \{k_i\}_{\rm in}$. The same conclusion is shown to hold when there are ingoing soft lines with momenta of the order of $k_{\rm out}$ in addition to the hard lines. It then becomes clear that the softness property is preserved if counterterms of the form $\d_j\tau^{ij}$ are added to equation \eqref{momentum}, where $\tau^{ij}$ is made of local observables. This ensures locality and momentum conservation.

%%%%%%%%%%%%%%%%%%%%%%%%%%%%%%%%%%%%%%%%%%%%%%%%%%%%%%%%%%%%
\section{A practical remedy}\label{sec:remedy}

So far we learned that a systematic renormalization of the two field $(\delta,\theta)$ formulation of cosmic fluid is possible if one uses nonlocal in time counterterms and add corrections to both continuity and Euler equations. However, there are three reasons to seek an alternative approach:

\begin{enumerate}[I]

\item One may not be interested in the velocity statistics, and even if she is, there are various ways to define velocity field and the one used in section \ref{counter} might not agree with the velocity field obtained from an actual survey or simulation.  

\item More importantly, the momentum conservation constraint is not straightforwardly implemented. 

\item The nonlocal in time counterterms introduce additional complications, since instead of a single time-dependent coefficient one needs to consider a multi-variable function of insertion times for every new counterterm.

\end{enumerate}

We first show in section \ref{velind} that one can use the momentum formulation to eliminate velocity. In this language, only the Euler equation receives corrections and not the continuity equation. Also, now there exists a straightforward prescription to generate all counterterms and only the ones allowed by the symmetries, including the conservation of total momentum. The observed velocity on the other hand must be treated similar to biased tracers; it can differ from the one calculated in any perturbative scheme by a set of locally observable counterterms whose coefficients must be empirically determined. This resolves the first two problems. Then in section \ref{sec:loc} we show that all nonlocal in time counterterms can be reduced perturbatively to a set of local operators, addressing problem III. This allows us to give a complete and more manageable list of counterterms.

%%%%%%%%%%%%%%%%%%%%%%%%%%%%%%%%%%%%%%%%%%%%%%%%%%

\subsection{Formulation in terms of the mass-weighted velocity}\label{velind}

If we use the density $\delta$ and momentum $\pib$ as the primary variables, the continuity equation reduces to the linear equation \eqref{cont}, and does not require any counterterm (it is not modified by smoothing) \cite{Carrasco:2012cv,2loop}. If the correlators of $ \delta $ (and therefore of $ \dot \delta $) are renormalized, then so are the correlators of $ \nabla \cdot \pib $ and vice versa. The same can be said for $ \phi $ and $ \delta $, given that the Poisson equation is also linear. The problem then reduces to solving the only non-linear equation, namely the vector Euler equation \eqref{momentum}. If one is interested in the correlators for $ \delta $, it is convenient to decompose the momentum into a divergence ($\mu\equiv\nabla\cdot\pib  $) and a vorticity ($ \nub\equiv \nabla\times\pib $), as discussed in appendix \ref{app:soft} (see \eqref{mu}, \eqref{delta1} and \eqref{nu}, see also \cite{Pueblas:2008uv}) . The Fourier space version of these equations is collected in appendix \ref{a:f}. Notice that all nonlinear terms on the right hand side of the Euler equation \eqref{momentum} are total derivatives of some tensor that is not proportional to $ \delta_{ij} $ and hence all contribute to the curl equation. In particular, all non-trivial solutions have $ \nub,\delta\neq 0 $. Using the continuity equation \eqref{cont}, one can get rid of $ \mu $ and solve the remaining equations for $ \delta $ and $ \nub $. When using momentum, one can therefore compute all correlators of $ \delta $ without including any counterterm in the continuity equation. We will see now that this is still the case when using a specific definition of the velocity.

In Standard Perturbation Theory (SPT), where $ \tau_{ij}=0 $, there exist a convenient change of variables for which the curl of the vector Euler equation is trivially satisfied even for non-trivial solutions $ \delta\neq0 $. From \eqref{momentum}, it is clear that a necessary condition for this to happen is that the curl is not sourced by the term $ (1+\delta) \partial_{i}\phi $. Upon dividing the whole equation \eqref{momentum} by $ \left( 1+\delta \right) $, this term reduces to the gradient of a scalar which disappears after taking the curl. This suggest to use the variable $ \v_{\pi}\equiv \pib/(1+\delta) $, which, as it is well-known, indeed decouples the vorticity $ \nabla \times \v_{\pi} $ from $ \delta $ in SPT. We will refer to $ \v_{\pi} $ as ``mass-weighted'' velocity.

More explicitly, let us define the divergence and curl of the velocity (we raise and lower all indices with $\delta_{ij}$)
\be
\theta_\pi\equiv \nabla\cdot \vec v_\pi=\partial_{i} v_\pi^{i}\,,\quad w_\pi^{i}\equiv\nabla \times \vec v_\pi=\epsilon^{ijk}\partial_{j} v_\pi^{k}\,,
\ee
from which it follows $\partial_{i} w_\pi^{i}=0$. The equations of motion then take the schematic form (see appendix \ref{a:f} for the explicit form)
\ba
\partial_{\tau} \delta+ \theta_\pi &=& -\alpha  \theta_\pi\delta + \va \cdot \vec w_\pi \delta\,,\label{c2}\\
\partial_{\tau} \theta_\pi+ \cH \theta_\pi+\frac{3}{2} \cH^{2}\Omega_{m} \delta&=& - \beta  \theta_\pi \theta_\pi + \vb \cdot\vec w_\pi\theta_\pi +\mb_{ij} w_\pi^{i}   w_\pi^{j}+\partial \cdot \left( \frac{\vec {\partial \tau} }{1+\delta}\right) \\
\partial_{\tau} \vec w_\pi +\cH \vec w_\pi&=& \mg_{ij}  w_\pi^{j} \theta_\pi +\va  \vec w_\pi \cdot  \vec w_\pi +\vg_{ijl}  w_\pi^{l}  w_\pi^{j}+\partial \times \left( \frac{\vec {\partial \tau} }{1+\delta}\right) \,,\label{c3}
\ea

%The higher order terms needed in this equation in the $(\delta,\theta)$ formalism of section \ref{counter} (or of \cite{Mercolli:2013bsa}) can be attributed to the fact that starting from
%\be\label{vL}
%\v_{\pi,\Lambda}=\frac{[\pib]_\Lambda}{1+[\delta]_\Lambda}\,,
%\ee
%where now we have made explicit the smoothing on a some scale $\Lambda$. Then, smoothing over a range of momenta to reach a lower scale $\Lambda'<\Lambda$, would result in a velocity field that differs from $\v_{\pi,\Lambda'}$ by the addition of some local counterterms. Thus working with $\v_\pi$ amounts to appropriate redefinitions as short modes are integrated out.

We can now perturbatively solve for $\theta_\pi $ and $ \w_\pi $ in terms of $\delta$ using \eqref{c2} and \eqref{c3}. This yield, order by order, a single differential equation for $ \delta $. Let us see explicitly how this works to quadratic order. The vorticity is sourced by $ \partial\times \left[ \partial\tau/(1+\delta) \right] $ and therefore starts quadratic in $ \delta $ perturbations. We can hence neglect it in the following formulae, keeping in mind that it will contribute to $ \delta $ at the next order, $ \mathcal{O}(\delta_{1}^{3}) $. We now solve the continuity equation 
\be\label{theta}
\theta_\pi = -\dot\delta + \d_i (\delta \frac{\d_i}{\nabla^2}\dot \delta) + \O(\delta^3)\,,
\ee
and substitute it into the divergence of the Euler equation to obtain a nonlinear equation purely in terms of $\delta$:
\be
\label{sptdelta}
\ddot\delta+\cH \dot\delta-\frac{3}{2}\cH^2 \delta=(\d_\tau+\cH)\d_i(\delta \frac{\d_i}{\nabla^2}\dot \delta)
+\d_i(\frac{\d_j}{\nabla^2}\dot \delta \frac{\d_i\d_j}{\nabla^2}\dot \delta)+\O(\delta^3).
\ee
The linear part of this equation is a second order differential equation which gives a retarded Green's function, using which one can perturbatively solve for the evolution of $\delta(\tau,\x)$ in terms of the initial condition $\delta_1(\x)$, without ever mentioning $\theta_\pi$. Obviously the resulting equation for $\delta$ now has interactions of all orders in $\delta$. On the other hand, the original system \eqref{sptcoupled} contains only cubic interactions and quadratic mixings between intermediate $\theta$ and $\delta$ lines. What the above procedure of solving for $\theta_\pi$ in terms of $\delta$ means is that all intermediate $\theta$ lines are contracted to points, thereby generating arbitrarily high order interactions of $\delta$. 

This formulation justifies a common approach in the EFT literature which is to use $\delta$ and $\theta$ as variables but to consider counterterms \textit{only} in the Euler equation while keeping the continuity equation intact. It is equivalent to renormalizing the equation \eqref{sptdelta}, as any conventional nonlinear theory of a single field $ \delta $, and to redefine the composite operator $\theta_\pi$ to absorb the counterterm $\nabla\cdot \bsb j_c$ of the continuity equation, so that the redefined $\theta_\pi$ is given by the same relation \eqref{theta} at all scales. However, doing so can in principle introduce vorticity because $\nabla \cdot \bsb j_c$ is not necessarily Laplacian of a locally observable scalar. Therefore, the counterterms of the (vectorial) Euler equation needed to renormalize the SPT loops are no longer pure gradients. This was the case in \cite{Baldauf}, where the quadratic counterterms $\d_j(\d_i\d_k\phi \d_j\d_k\phi)$, which is not the pure gradient of a locally observable scalar, was needed to renormalize one-loop bispectrum of $\delta$. 

In summary, one can use the standard perturbation theory approach to calculate various correlation functions of $\delta$. The counterterms that are allowed to be added to the vector Euler equation are now fixed by momentum conservation to be of the form 
\be\label{ddtau}
(1+\delta)^{-1}\d_j\tau^{ij}
\ee
where in order to do systematic renormalization $\tau^{ij}$ must be allowed to depend on observable quantities along the flow, and hence be nonlocal in time.

As for the statistics of the observed velocity field, the difference with that of perturbation theory (say $\v_\pi$) can be parameterized by a set of counterterms
\be
\v_{\rm obs} = \v_{\pi}+\v_{\rm ct},
\ee
where $\v_{\rm ct}$ is a locally observable vector. It has a non-stochastic piece which by symmetry must start from first order in derivative counting, and a stochastic piece which according to the arguments of appendix \ref{app:v} is generically $\O(k^0)$, leading to $\O(k^2)$ contribution to the vorticity power spectrum. Hence, the treatment of velocity is similar to the biased tracers. In particular the most general $\v_{\rm ct}$ can be expressed locally in time \cite{bias}.

%%%%%%%%%%%%%%%%%%%%%%%%%%%%%%%%%%%%%%%%%%%%%%%%%%%%%%%%%%%
\subsection{Local formulation\label{sec:loc}}

It was argued in the last section that a generic counterterm can be constructed out of a tensor $\tau^{ij}$. The tensor is made of observable quantities integrated along the flow:
\be\label{ct2}
\tau^{ij}(\x,\tau)=
\int^\tau_{0}\{d\tau_n\} K_{a\cdots a_n\cdots}(\tau;\{\tau_n\}) \cdots \Pi_{a_n}^{i_n j_n}(\xfl(\tau_n),\tau_n)\cdots,
\ee
where we switched to the conformal time $\tau$. In perturbation theory, any observable can be solved in powers of the initial field $\delta_1(\x)$, in particular,
\be
\Pi(\x,\tau) = D(\tau) \Pi^{(1)}(\x)+D^2(\tau)\Pi^{(2)}(\x)+\cdots
\ee
where $D(\tau)$ is the growth factor and we are using the accurate approximation of replacing $a(\tau)\to D(\tau)$ in going from matter dominance to $\Lambda$CDM. This expansion contains displacement terms to take into account the bulk flow induced by long wavelength initial perturbations. As such, some of the second and higher order terms in the sum become singular in the presence of a finite $\delta_1(\k_1)$ with vanishing momentum $k_1\to 0$. These terms are responsible for shifting the argument of the fields on the right hand side (r.h.s.) to their initial (Lagrangian) fluid position
\be
\z = \xfl[0;\x,\tau].
\ee
After shifting the arguments $\x\to \z$ there remains no infra-red singularity on the r.h.s. as expected on symmetry grounds. We denote the operators thus obtained Lagrangian operators. In terms of them \cite{Baldauf,bias}
\be
\Pi(\x,\tau) = D(\tau) \Pi_\lgr^{(1)}(\z)+D^2(\tau)\Pi_\lgr^{(2)}(\z)+\cdots
\ee
Substituting this expansion in \eqref{ct2}, makes it possible to perform all time integrations to obtain an expansion of various products of Lagrangian quantities with coefficients that generically depend on (and only on) the final time $\tau$. The whole expression is acted on by two Eulerian derivatives:
\be\label{ca}
\frac{\d}{\d x^i}\left[\frac{1}{1+\delta}\frac{\d}{\d x^j}
\sum_{\{m_n\}}k(\tau;\{m_n,a_n\})\cdots [\Pi_{a_n}^{i_n j_n}]_\lgr^{(m_n)}(\z)\cdots\right]
\ee
It was shown in \cite{bias} that there is a one-to-one correspondence between the Lagrangian quantities and the locally measurable (Eulerian) operators. Hence, we can re-express \eqref{ca} in terms of those Eulerian operators. A basis for the latter can be obtained by combining various convective time derivatives \eqref{convective} of $\d_i\d_j\phi$. Starting from
\be
\hat\Pi^{(1)}_{ij}(\x,\tau) =\: \d_i\d_j\phi(\x,\tau),
\ee
the convective derivatives can be combined in such a way that a set of operators with increasing order in perturbation theory is obtained. Suppressing the indices and the argument $(\x,\tau)$ these higher order operators can be computed recursively:
\be
\hat\Pi^{(n)} =\:\frac{1}{(n-1)!} \left[(\hat\Pi^{(n-1)})' - (n-1)\hat\Pi^{(n-1)}\right],
\ee
where prime denotes convective derivative with respect to $\log D$ ($=\eta$ in a matter dominated universe). At leading order these operators match those that appear in the perturbative expansion of $\d_i\d_j\phi$:
\be
\Pih^{(n)}(\x,\tau)= D^n(\tau) \Pi_\lgr^{(n)}(\z)+\cdots
\ee
At lowest derivative level an Eulerian basis for two-index tensors can be constructed as follows. First, a two index tensor is made out of a (matrix) product of a (possibly empty) sequence of $\Pih^{(n)}$'s. For instance, $\delta_{ij}$, $\Pih^{(1)}_{ik}\Pih^{(3)}_{kl}\Pih^{(1)}_{lj}$ are two examples. This tensor can be multiplied by an arbitrary scalar constructed out of traces of products of $\hat\Pi^{(n)}_{ij}$, like $\Pih^{(1)}_{ij}\Pih^{(1)}_{ij}$. The trace $\tr[\Pih^{(n)}]$ with $n>1$ must be excluded since it is expressible in terms of lower order operators. The first few orders are (in matrix notation):
\ba
{\rm 1^{st}} \ && \ \unit {\rm Tr}[\Pih^{(1)}],\Pih^{(1)}  \label{listP} \\ 
{\rm 2^{nd}} \ && \ \unit {\rm Tr}[(\Pih^{(1)})^2],\  \unit ({\rm Tr}[\Pih^{(1)}])^2,\ \Pih^{(1)} \tr[\Pih^{(1)}],\
(\Pih^{(1)})^2,\ \Pih^{(2)} \nonumber\\ 
{\rm 3^{rd}} \ && \ \unit {\rm Tr}[(\Pih^{(1)})^3 ],\ \unit {\rm Tr}[(\Pih^{(1)})^2 ]  {\rm Tr}[\Pih^{(1)}],\ \unit ({\rm Tr}[\Pih^{(1)}])^3,\ \unit {\rm Tr}[\Pih^{(1)} \Pih^{(2)}],\cdots \nonumber 
\ea
Note that the $\tau^{ij}$ matrix must be symmetrized. To obtain a list of counterterms, one substitutes any such tensor into the $ \tau_{ij} $ of \eqref{ddtau}. Let us close with a few clarifications.

\begin{itemize}

\item This is a complete basis at lowest derivative level. In particular, all locally observable operators made of the velocity field are also included. For instance, in terms of a rescaled velocity potential $\nabla^2 \phi_v=-\cH^{-1}\theta$, we have
\be
\hat\Pi^{(2)}_{ij} = \d_i\d_m\phi\d_j\d_m\phi+\frac{5}{2}(\d_i\d_j\phi-\d_i\d_j\phi_v)+\O(\delta^3).
\ee
Clearly, at higher orders $\d_i\d_j\phi$ and $\d_i\d_j\phi_v$ will be insufficient for a local description of counterterms and higher order convective time derivatives will be necessary. However, given that at $n^{th}$ order we need up to $\Pih^{(n-1)}$, the first example appears at the quartic level. 

\item The tensors $\hat \Pi^{(n)}(\x,\tau)$, and hence the counterterms in \eqref{ddtau}, are easily calculable in momentum space by combining the $F$ and $G$ kernels of SPT. For instance,
\be
\hat\Pi^{(1)}_{ij}(\k,\tau) =\:\frac{\k^i \k^j}{|\k|^2}\sum_{n=1}^{\infty}D^n(\tau) \delta_n(\k),
\ee
where $\delta_n(\k)$ is the $n^{th}$ order SPT solution, and 
\be
\begin{split}
\hat\Pi^{(2)}_{ij}(\k,\tau) =\:\sum_{n=1}^{\infty}D^n(\tau)
\Big[&\frac{\k^i \k^j}{|\k|^2} (n-1)\delta_n(\k)\\[8pt]
&+\sum_{m=1}^{n-1}\int_\p\frac{(\k-\p)\cdot \p}{|\k-\p|^2}\frac{\p^i \p^j}{|\p|^2}\theta_m(\k-\p)\delta_{n-m}(\p)\Big],
\end{split}
\ee
and so on.

\item One should bear in mind that although $\d_i\d_j\phi(\x,\tau)$ is a total derivative, neither its convective time derivatives nor the $\z$-space matrices $\Pi_{ij}^{(n)}(\z)$ are necessarily total spatial derivatives of any kind.

\item Although, the basis \eqref{listP} for tensors is non-redundant, taking the derivatives leads to degeneracies. For instance, at second order there are 5 tensors but only 3 indepedent counterterms. 

\item Any tensor $\tau^{ij}$ leads to an infinite series of counterterms when the factor $(1+\delta)^{-1}$ is expanded in \eqref{ddtau}. The relative coefficients of these counterterms are fixed so as to insure double softness of $\delta$ when the ingoing fields are hard (a fact that is verified in appendix \ref{app:soft} using the momentum equations).

\end{itemize}

%%%%%%%%%%%%%%%%%%%%%%%%%%%%%%%%%%%%%%%%%%%

\section{Conclusions}

In this paper, we have introduced a systematic renormalization procedure for the perturbative description of Large Scale Structures, which is analogous to the one commonly used in quantum field theory. One main advantage is that new counterterms can be straightforwardly read off amputated time-evolution diagrams, without the complicated mixing with lower-order counterterms that plagues the standard treatment. In this approach, correlators of both $ \delta $ and $ \theta $ are renormalized at each order. We have also provided for the first time an explicit prescription for the allowed counterterms to all orders in perturbation theory. In passing, we have justified the commonly used mass-weighted volcity formulation in which counterterms appear exclusively in the Euler equation.\footnote{This had been checked so far explicitly only for the one-loop power spectrum \cite{Mercolli:2013bsa} and at linear order in the leading counterterm to the renormalized velocity \cite{2loop}.} We have also presented a rigorous derivation of the double-softness of $ \delta $ and $ \nabla \cdot \pib $ in perturbation theory, namely the well-known fact that $ \delta(\k)\sim \nabla \pib(\k) \sim k^{2} $ on large scales, $ k\rightarrow 0 $.

%%%%%%%%%%%%%%%%%%%%%%%%%%%%%%%%%%%%%%%%%%%%%%
\section*{Acknowledgments}

We thank Valentin Assassi, Tobias Baldauf, Daniel Green, Lorenzo Mercolli, Marko Simonovi\'c, Drian van der Woude, Yvette Welling for useful discussions and comments on the manuscript. Special thanks go to Matias Zaldarriaga for his invaluable insights and inputs throughout the project. MM acknowledges support from NSF Grants PHY-1314311, PHY-0855425, and PHYS-1066293, and is grateful for the hospitality of the Aspen Center for Physics. The work of MP is supported in part by NSF grant PHY-1316452. E.P. is supported by the Delta- ITP consortium, a program of the Netherlands organization for scientific research (NWO) that is funded by the Dutch Ministry of Education, Culture and Science (OCW). A.A.A. is partially supported by Deputy of Research of SUT.

%%%%%%%%%%%%%%%%%%%%%%%%%%%%%%%%%%%%%%%%%%%%%%%
\appendix

%%%%%%%%%%%%%%%%%%%%%%%%%%%%%%%%%%%%%%%%%%%%%%%%%%

\section{Second order counter terms at 1-loop order} 

\begin{figure}
    \begin{subfigure}[b]{0.42\textwidth}
        \centering
        \resizebox{\linewidth}{!}{
            \begin{tikzpicture}
\draw [very  thick](.1,0) -- (1,0);
\draw [very  thick] (1,0) -- (3,2);
\draw [very  thick](1,0) -- (3,-2);
\draw [very  thick](3,-2) -- (3,2);
\draw [very  thick](3,2) -- (4,3);
\draw [very  thick](3,-2) -- (4,-3);

\draw (.4,0) node {$\blacktriangleleft$};
\draw (3.4,2.4) node[rotate=45] {$\blacktriangleleft$};
\draw (3.4,-2.4) node[rotate=-45] {$\blacktriangleleft$};
\draw (2,-1) node[rotate=-45] {$\blacktriangleleft$};
\draw (2,1) node[rotate=45] {$\blacktriangleleft$};
\draw (3,1) node[rotate=90] {$\blacktriangleright$};
\draw (3,-1) node[rotate=90] {$\blacktriangleleft$};

\draw (3,0) node {\scalebox{1.5}{$\bullet$}};

\draw (-.5,0) node {$a, \K$};
\draw (1,-.5) node {$\eta$};
%\draw (.5,-.5) node{$\K$};
\draw (4.7,3.2) node {$a_1, \K_1$};
%\draw (3.8,2) node{$\K_1$};
\draw (2.8,2.5) node{$\eta_1$};
\draw (4.7,-3.2) node {$a_2,\K_2$};
%\draw (3.8,-2) node{$\K_2$};
\draw (2.8,-2.5) node{$\eta_2$};
\draw (7,.5) node{$ \K_1 \leftrightarrow \K_2$};
\draw (7,0) node{$\eta_1 \leftrightarrow \eta_2$};
\draw (7,-.5) node{$a_1 \leftrightarrow a_2$};
\draw (5.75,0) node{$+\Bigg ( $};
\draw (8,0) node{$\Bigg )$};

\end{tikzpicture}
        }
        \caption{Amputated Diagram.1}
        \label{amp-dia1}
    \end{subfigure}
    \qquad
    \begin{subfigure}[b]{0.42\textwidth}
    \centering
        \resizebox{\linewidth}{!}{
            \begin{tikzpicture}
\draw [very  thick](.1,0) -- (1,0);
\draw [very  thick](1,0) -- (3,2);
\draw [very  thick](1,0) -- (3,-2);
\draw [very  thick](3,2) -- (3,-2);
\draw [very  thick](3,2) -- (4,3);
\draw [very  thick](3,-2) -- (4,-3);

\draw (.4,0) node {$\blacktriangleleft$};
\draw (3.4,2.4) node[rotate=45] {$\blacktriangleleft$};
\draw (3.4,-2.4) node[rotate=-45] {$\blacktriangleleft$};
\draw (1.5,.5) node[rotate=45] {$\blacktriangleleft$};
\draw (2.5,1.5) node[rotate=45] {$\blacktriangleright$};
\draw (2,-1) node[rotate=-45] {$\blacktriangleleft$};
\draw (3,0) node[rotate=90] {$\blacktriangleleft$};

\draw (2,1) node {\scalebox{1.5}{$\bullet$}};

\draw (-.5,0) node {$a,\K$};
%\draw (.5,-.5) node{$\K$};
\draw (1,-.5) node {$\eta$};
\draw (4.7,3.2) node {$a_1,\K_1$};
%\draw (3.8,2) node{$\K_1$};
\draw (2.8,2.5) node{$\eta_1$};
\draw (4.7,-3.2) node {$a_2,\K_2$};
%\draw (3.8,-2) node{$\K_2$};
\draw (2.8,-2.5) node{$\eta_2$};
\draw (7,.5) node{$ \K_1 \leftrightarrow \K_2$};
\draw (7,0) node{$\eta_1 \leftrightarrow \eta_2$};
\draw (7,-.5) node{$a_1 \leftrightarrow a_2$};
\draw (5.75,0) node{$+\Bigg ( $};
\draw (8,0) node{$\Bigg )$};
\end{tikzpicture}
        }
        \caption{Amputated Diagram.2}   
        \label{amp-dia2}
    \end{subfigure}
\caption{1-loop corrections to the cubic vertex function}
\label{fig:subfig1.a.4}
\end{figure}
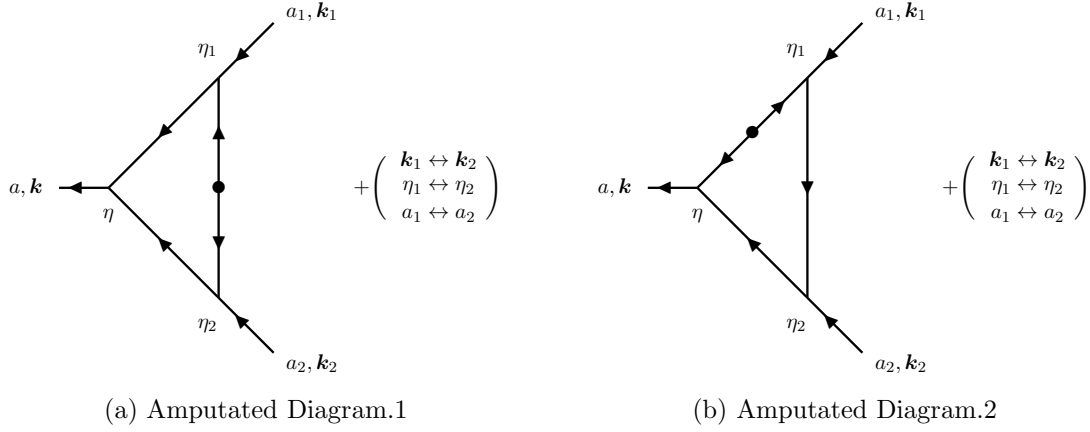

This section is devoted to a detailed calculation of second order counterterms in the leading loop order. This is sufficient for a 1-loop calculation of the bispectrum. However, the resulting counterterms renormalize all reducible diagrams where the vertex corrections of Fig. \ref{amp-dia1} and \ref{amp-dia2} appear as a sub-diagram, which is the goal of systematic renormalization. For this the counterterms are unavoidably nonlocal in time since the ingoing lines into this sub-diagram can be of different orders and hence have different time-dependence. This is the practical manifestation of the ``long memory effect'' discussed in section \ref{memory}.

 Using Feynman rules presented in section \ref{pre}, the 1-loop correction to the vertex function can be read off diagrams of Fig.\ref{amp-dia1} and Fig.\ref{amp-dia2} as

\be
\begin{split}
C^{\mathrm{1-loop}}_{a,a_1,a_2}(\eta,\eta_1,\eta_2;\k,\k_1,\k_2) =~ C^{\mathrm{dia.1}}_{a,a_1,a_2}(\eta,\eta_1,\eta_2;\k,\k_1,\k_2)&+ C^{\mathrm{dia.2}}_{a,a_1,a_2}(\eta,\eta_1,\eta_2;\k,\k_1,\k_2) \\[10pt]
&+\Bigg(
\begin{array}{lr}
 \k_1 \leftrightarrow \k_2
\\
\,\eta_1 \leftrightarrow \eta_2
\\
\,a_1 \leftrightarrow a_2
\end{array}
\Bigg )
\end{split}
\ee
for which the diagram.1 and diagram.2 can be calculated through performing hard momentum (loop momentum) integrations of the following equations. At this step time integrals are left untouched  
\begin{align}
\nonumber
 C^{\mathrm{\mathrm{dia.}1}}_{a\,a_1\,a_2}(\eta,\eta_1,\eta_2;\k,\k_1,\k_2) &=
4 \times g_{f\,m}(\eta_2,0) g_{h\,l}(\eta_1,0) g_{b\,e}(\eta,\eta_1) g_{c\,d}(\eta,\eta_2) 
\\
\label{deltaC-dia1}
& \int_{\Q}\gamma _{a,b,c}\left(\K ,\K_1 +\Q ,\K_2 - \Q\right) \gamma _{d,f,a_2}\left(\K_2- \Q,- \Q, \K_2\right) \gamma _{e\,a_1\,h}\left(\K_1 +\Q, \K_1, \Q\right)
\\
\nonumber
C^{\mathrm{\mathrm{dia.}2}}_{a\,a_1\,a_2}(\eta,\eta_1,\eta_2;\k,\k_1,\k_2) &=
8 \times g_{b\,l}(\eta,0) g_{e,m}(\eta_1,0) g_{c\,d}(\eta,\eta_2) g_{f\,h}(\eta_2,\eta_1)
\\
\label{deltaC-dia2}
& \int_{\Q} \gamma _{a\,b\,c}\left( \K ,-\Q, \K+ \Q \right) \gamma _{d\,f\,a_2}\left(\K+ \Q, \K_1+ \Q, \K_2\right) \gamma _{h\,e\,a_1}\left( \K_1+\Q,\Q,\K_1 \right)
\end{align}
and summation on repeated indices is presumed. It should be noticed that the counterpart of diagram.1 which can be found by exchanging $ \k_1 \leftrightarrow \k_2 , \eta_1 \leftrightarrow \eta_2,a_1 \leftrightarrow a_2$, would just multiply the contribution of diagram.1 in $C^{\rm 1-loop}_{a \,a_1\,a_2}$ by a factor $2$. However, in the second diagram, $(1\leftrightarrow 2)$ conversion produces a new momentum and time dependence.

% diagram acting on the second diagram which diagrammatically means putting bullet (indicating initial field contractions) on the $\K_3$ line (instead on $\K_2$ line in the \ref{amp-dia2}). 

%%%%%%%%%%%%%%%%%%%%%%%%%%%%%%%%%%%%%%%%%%%%%%%%%%%
%%%%%%%%%%%%%%%%%%%%%%%%%%%%%%%%%%%%%%%%%%%%%%%%%%%

The results of the above integrals are 
\begin{align}
\label{C-dia1}
\nonumber
 C^{\mathrm{dia.1}}_{a,a_1,a_2}(\k;\k_1,\k_2; \eta,\eta_1,\eta_2)=~& 
 \sigma^2~\theta(\eta-\eta_1)~\theta(\eta-\eta_2)~
 \\
 \nonumber
 \Bigg \{
&{}^1\Gamma^{(1)}_{a,a_1,a_2} ~e^{2\,\eta}+{}^1\Gamma^{(2)}_{a,a_1,a_2} ~e^{(1/2)\,(-\eta+5\eta_1)}+
\\
&{}^1\Gamma^{(3)}_{a,a_1,a_2} ~e^{(1/2)\,(-\eta+5\eta_2)}+    {}^1\Gamma^{(4)}_{a,a_1,a_2} ~e^{(5/2)\,(\eta_1+\eta_2)-3\eta}
\Bigg \}
\end{align}
\begin{align}
\label{C-dia2}
\nonumber
 C^{\mathrm{dia.2}}_{a,a_1,a_2}(\k;\k_1,\k_2; \eta,\eta_1,\eta_2)=~& 
 \sigma^2~\theta(\eta-\eta_2)~\theta(\eta_2-\eta_1)~
\\ 
\nonumber
 \Bigg \{&
{}^2\Gamma^{(1)}_{a,a_1,a_2} ~e^{2\,\eta}+{}^2\Gamma^{(2)}_{a,a_1,a_2} ~e^{(1/2)\,(-\eta+5\eta_1)} +     
\\
&{}^2\Gamma^{(3)}_{a,a_1,a_2} ~e^{(1/2)\,(-\eta+5\eta_2)}+    {}^2\Gamma^{(4)}_{a,a_1,a_2} ~e^{(5/2)\,(\eta_1-\eta_2)+2\eta}
\Bigg \}
\end{align}
in which ${}^1\Gamma^{(n)}_{a,b,c}$ and ${}^2\Gamma^{(n)}_{a,b,c}$  are originating from 1-loop vertex  diagram $1$ and diagram $2$ respectively.

%%%%%%%%%%%%%%%%%%%%%%%%%%%%%%%%%%%%%%%%%%%%%%%%%%%%%%%%%%%%%%%%%%%
\subsection{Shift terms \label{app:IR}}

As discussed in section \ref{memory} promoting the time integration to be along the flow identifies several higher order counterterms as the displacement a lower order one. These shift terms can be isolated by looking at the infra-red limit when one of the ingoing momenta becomes much smaller than the other. In other words, the two vertices $v^i \partial_i \theta$ and $v^i \partial_i \delta$ with soft velocities lead to IR-divergences as $1/k$ in the momentum of $\Psi_2 \sim \theta = \nabla.v$. In the case of 1-loop second-order counterterms, the first order shift of a first order counterterm \eqref{shift} automatically renormalize these ``IR-divergent'' cutoff dependence. It is shown in section \ref{1PI-shift-sec} that these non-original counterterms are
\ba
\nabla^2 \int^{\eta}_{\eta_0} d\eta_1 \int^{\eta}_{\eta_1} d\eta_2 ~c_{a\,b}(\eta,\eta_1) ~\dfrac{\nabla}{\nabla^2} 
\Psi_2(\x,\eta_2).\nabla\,\Psi_{b}(\x,\eta_1).
\ea
Now using the 1-loop correction of the propagator
\ba
\label{delta-propag-amp-2}
\nonumber
c_{a\,b}(\eta,\eta_1)&=& - \sigma^2_v ~~ e^{-\eta/2}\Bigg\{
\Bigg[  \begin{array}{ccc}
\frac{3}{5} & \frac{-4}{25}
\\
\frac{-9}{5} & \frac{-8}{5}
\end{array}
\Bigg]~  e^{\frac{5 \eta }{2} }+
%%%
\Bigg[  \begin{array}{ccc}
\frac{-8}{5} & \frac{4}{25}
\\
\frac{9}{5} & \frac{3}{5}
\end{array}
\Bigg] ~e^{\frac{5 \eta_1 }{2}} \Bigg\} ,
\ea 
the above correction to the equation of motion can be written in Fourier space as
\ba
\int^{\eta}_{\eta_0} d\eta_1 \int^{\eta}_{\eta_0} d\eta_2 ~\tilde{C} _{aa_1a_2} (\eta,\eta_1,\eta_2;\k,\k_1,\k_2) ~\Psi_{b}(\K_1,\eta_1)~\Psi_2(\K_2,\eta_2).
\ea
in which
\begin{align}
\label{vert-cor-dmflow}
\nonumber
\tilde{C} _{aa_1a_2} (\eta,\eta_1,\eta_2;\k,\k_1,\k_2) = 
 - \sigma^2~\delta _{a_2 2}~ \dfrac{k^2 \,(\k_1.\k_2)}{k^2_2}  &~\int^{\eta}_{\eta_0} d\eta_1 \int^{\eta}_{\eta_0} d\eta_2 ~\theta (\eta-\eta_1)~\theta (\eta_1-\eta_2)~
\\
~&\Bigg\{
\Bigg[  \begin{array}{ccc}
\frac{3}{5} & \frac{-4}{25}
\\
\frac{-9}{5} & \frac{-8}{5}
\end{array}
\Bigg]_{a\,a_1}~  e^{2\, \eta  }+
%%%
\Bigg[  \begin{array}{ccc}
\frac{-8}{5} & \frac{4}{25}
\\
\frac{9}{5} & \frac{3}{5}
\end{array}
\Bigg]_{a\,a_1} ~e^{\frac{5 \eta_1-\eta }{2}} \Bigg\}
\end{align}
Now let us examine the IR-limit of the corrections to the vertex function. Using the following identity between step functions
\ba 
\theta (\eta-\eta_1)\,\theta (\eta-\eta_2) = \theta (\eta-\eta_1)\,\theta (\eta_1-\eta_2)+\theta (\eta-\eta_2)\,\theta (\eta_2-\eta_1)
\ea
one readily finds that IR-limit of diagram.1 contribution to the vertex correction as
\begin{align}
\nonumber
\lim_{k_2 \rightarrow 0} C^{dia. 1} _{aa_1a_2} (\k,\k_1,\k_2;\eta,\eta_1,\eta_2) +(1 \leftrightarrow 2)&  =
 -\,2\times \sigma^2~\delta _{a_2\,2}~ \dfrac{k^2 \,(\k_1.\k_2)}{k^2_2} \int^{\eta}_{\eta'} d\eta_1 \int^{\eta}_{\eta'} d\eta_2  
\\ 
\nonumber
  ~& \Big[ \theta (\eta-\eta_1)\,\theta (\eta_1-\eta_2)+\theta (\eta-\eta_2)\,\theta (\eta_2-\eta_1)\Big]~\times
\\
&\Bigg\{
\Bigg[  \begin{array}{ccc}
\frac{6}{25} & \frac{12}{175}
\\
0 &  \frac{-6}{25}
\end{array}
\Bigg]_{a\,a_1}~  e^{2\, \eta  }+
%%%
\Bigg[  \begin{array}{ccc}
\frac{-6}{25} & \frac{-12}{175}
\\
0 & \frac{6}{25}
\end{array}
\Bigg]_{a\,a_1} ~e^{\frac{5 \eta_1-\eta }{2}} \Bigg\}
\end{align}
where the symmetry factor $2$ comes from $(1 \leftrightarrow2)$ or more explicitly $(\k_1 \leftrightarrow \k_2,\eta_1 \leftrightarrow \eta_2,a_1 \leftrightarrow a_2) $. And 
\begin{align}
\nonumber
\lim_{k_2 \rightarrow 0}\delta C^{dia. 2} _{aa_1a_2} (\k,\k_1,\k_2 &; \eta,\eta_1,\eta_2)+(1 \leftrightarrow 2) = 
 -  \sigma^2~\delta _{a_2\,2}~   \dfrac{k^2 \,(\k_1.\k_2)}{k^2_2}  ~\int^{\eta}_{\eta_0} d\eta_1 \int^{\eta}_{\eta'} d\eta_2 ~ \Bigg [ 
\\
\nonumber
& \theta (\eta-\eta_2)~\theta (\eta_2-\eta_1)~\Bigg\{
\Bigg[  \begin{array}{ccc}
\frac{-12}{25} & \frac{-24}{175}
\\
0 &  \frac{12}{25}
\end{array}
\Bigg]_{a\,a_1}~  e^{2\, \eta  }+
%%%
\Bigg[  \begin{array}{ccc}
\frac{12}{25} & \frac{24}{175}
\\
0 & \frac{-12}{25}
\end{array}
\Bigg]_{a\,a_1} ~e^{\frac{5 \eta_1-\eta }{2}} \Bigg\}
\\
 -  & \theta (\eta-\eta_1)~\theta (\eta_1-\eta_2)~\Bigg\{
\Bigg[  \begin{array}{ccc}
~\frac{3}{25} & \frac{-52}{175}
\\
\frac{-9}{5} &  \frac{-28}{25}
\end{array}
\Bigg]_{a\,a_1}~  e^{2\, \eta  }+
%%%
\Bigg[  \begin{array}{ccc}
\frac{-28}{25} & \frac{52}{175}
\\
\frac{9}{5} & \frac{3}{25}
\end{array}
\Bigg]_{a\,a_1} ~e^{\frac{5 \eta_1-\eta }{2}} \Bigg\} \Bigg ].
\end{align}
Again the third line is the soft limit amplitude of the inverted version of Diagram.2. Adding up these contributions together one recover the IR-divergent vertex correction already anticipated by promoting time integrals in propagator correction integral, $C^{UV}_a$, to an integral along $\xfl$, which leads to \eqref{vert-cor-dmflow}. 

%%%%%%%%%%%%%%%%%%%%%%%%%%%%%%%%%%%%%%%%%%%%%%%%%%%%
\subsection{The new counterterms \label{app:2ndct}}

As discussed in \ref{1PI-new-ct}, after the subtraction of shift terms from 1PI diagrams, what remains must be renormalized with the  new higher order counterterms. For every second order term (counter-term) in the equation of motion there is an associated cubic vertex in Feynman diagrams so we use 1-loop second order counterterms and 1-loop cubic vertex correction interchangeably. Using Eq. \eqref{c2}, all second order counterterms are of the following form
\be
\int^\eta_{\eta_0} d\eta_1\int^\eta_{\eta_0} d\eta_2 ~K^i_{a a_1 a_2}(\eta,\eta_1,\eta_2)~F^i(\K,\K_1,\K_2)~
\Psi_{a_1}(\K_1,\eta_1)\Psi_{a_2}(\K_2,\eta_2).
\ee
The general structure of $F^i$ kernels can be found using  Eq. \eqref{2nd} as
\begin{align}
\nonumber
F^{1}(\K,\K_1,\K_2) &= \K.\K_1 \qquad ~~~~~~~~~~~~~~ F^{2}(\K,\K_1,\K_2)= (\K.\K_2)
\\
\nonumber
 F^{3}(\K,\K_1,\K_2)&= \dfrac{(\K.\K_1)\,(\K_1.\K_2)}{k_1^2}\qquad ~F^{4}(\K,\K_1,\K_2)= \dfrac{(\K.\K_2)\,(\K_1.\K_2)}{k_2^2}
\\
\tilde F^{5}(\K,\K_1,\K_2)&= \dfrac{(\K.\K_1)(\K_1.\K_2)^2}{k_1^2\,k_2^2}\qquad \tilde F^{6}(\K,\K_1,\K_2)= \dfrac{(\K.\K_2)(\K_1.\K_2)^2}{k_1^2\,k_2^2}
 \end{align}
As already emphasized, although the basis \ref{2nd} for tensors is non-redundant, taking the derivatives leads to degeneracies. It can be shown that the above structures are linearly dependent: $\tilde F^6-\tilde F^5=F^3-F^4$. Hence, we eliminate $\tilde F^6-\tilde F^5$ and define 
 \ba
{F}^5 =\dfrac{1}{2}(\tilde F^5+\tilde F^6).
 \ea
Having found all possible momentum dependencies, the next step would be listing the time kernels. There is a unique basis for all possible time kernels of 2-nd order counterterms in the equation of motion. We denote them by  $\hat{\kappa}_{i}$ given by
 \begin{align}
 \nonumber
 \hat{\kappa}_1 &= e^{2 \eta}
 &\hat{\kappa}_2 &= e^{(5\eta_1-\eta)/2}
 &\hat{\kappa}_3 &=  e^{(5\eta_2-\eta)/2}
 \\
 \hat{\kappa}_4 &= e^{5(\eta_1-\eta_2)/2+2 \eta}  &\hat{\kappa}_5 &=e^{5(\eta_1-\eta_2)/2+2 \eta}  &\hat{\kappa}_6 &= e^{5(\eta_1+\eta_2)/2-3 \eta}
 \end{align}
Every time kernel has three indices associated with indices of the one outgoing and two ingoing legs  respectively. As every index can be either $1$ or $2$ there are $2^3 = 8 $ distinct counter terms which are listed below
\begin{align}
\nonumber
\sum_i K^i_{121}F^i &=+\frac{1}{125} F^1 \big[-6  (4 \hat{\kappa}_1-4 \hat{\kappa}_2-7 \hat{\kappa}_5+7 \hat{\kappa}_6)\Theta _{12}- (24 \hat{\kappa}_1-4 \hat{\kappa}_3+63 \hat{\kappa}_4+42 \hat{\kappa}_6)\Theta _{21}\big]
\\
\nonumber
&+ \dfrac{1}{125} F^2 \big[3  (44 \hat{\kappa}_1+25 \hat{\kappa}_2-74 \hat{\kappa}_3+3 \hat{\kappa}_4+2 \hat{\kappa}_6) \Theta _{21}+ (132 \hat{\kappa}_1-57 \hat{\kappa}_2-200 \hat{\kappa}_3-6 \hat{\kappa}_5+ 6 \hat{\kappa}_6) \Theta _{12}\big]
\\
&-\frac{2}{125} F^3 \big[6  (13 \hat{\kappa}_1-13 \hat{\kappa}_2-4 \hat{\kappa}_5+4 \hat{\kappa}_6)\Theta _{12}+ (78 \hat{\kappa}_1-13 \hat{\kappa}_3+36 \hat{\kappa}_4+24 \hat{\kappa}_6)\Theta _{21}\big]
%%%%%%%%%%%%%
\\
\nonumber
\sum_i K^i_{222}F^i &= \frac{3}{125} 
(F^1+F^2) \big[ (-74 \hat{\kappa}_1+25 \hat{\kappa}_2+44 \hat{\kappa}_3+2 \hat{\kappa}_4+3 \hat{\kappa}_6)\Theta _{21} +
\\
\nonumber
&~~~~~~~~~~~~~~~~~~~~~~~~ (-74 \hat{\kappa}_1+44 \hat{\kappa}_2+25 \hat{\kappa}_3+2 \hat{\kappa}_5+3 \hat{\kappa}_6)\Theta _{12}\big]
\\
&-\frac{2}{125} F^5 \big[   (-13 \hat{\kappa}_1+78 \hat{\kappa}_2+24 \hat{\kappa}_5+36 \hat{\kappa}_6)\Theta _{12}+(-13 \hat{\kappa}_1+78 \hat{\kappa}_3+24 \hat{\kappa}_4+36 \hat{\kappa}_6)\Theta _{21}\big]
%%%%%%%%%%
\\
\nonumber
\sum_i K^i_{122}F^i &=+\frac{2}{875} F^1 \big[\Theta _{21} (129 \hat{\kappa}_1+70 \hat{\kappa}_2-199 \hat{\kappa}_3+33 \hat{\kappa}_4-33 \hat{\kappa}_6)+3 \Theta _{12} (43 \hat{\kappa}_1-43 \hat{\kappa}_2+11 \hat{\kappa}_5-11 \hat{\kappa}_6)\big]
\\
\nonumber
&+\dfrac{2}{875} F^2 \big[\Theta _{12} (129 \hat{\kappa}_1-199 \hat{\kappa}_2+70 \hat{\kappa}_3+33 \hat{\kappa}_5-33 \hat{\kappa}_6)+3 \Theta _{21} (43 \hat{\kappa}_1-43 \hat{\kappa}_3+11 \hat{\kappa}_4-11 \hat{\kappa}_6)\big]
\\
\nonumber
&+\frac{1}{875} (F^3+F^4+F^5) \big[\Theta _{12} (-611 \hat{\kappa}_1+611 \hat{\kappa}_2+48 \hat{\kappa}_5-48 \hat{\kappa}_6)
\\
&~~~~~~~~~~~~~~~~~~~~~~~~~~~+\Theta _{21} (-611 \hat{\kappa}_1+611 \hat{\kappa}_3+48 \hat{\kappa}_4-48 \hat{\kappa}_6)\big]
%%%%%%%%%
\\
\nonumber
\sum_i K^i_{111}F^i &= \frac{3}{25} F_1 \big[\Theta _{12} (6 \hat{\kappa}_1-16 \hat{\kappa}_2+15 \hat{\kappa}_3-3 \hat{\kappa}_5-2 \hat{\kappa}_6)+\Theta _{21} (6 \hat{\kappa}_1-\hat{\kappa}_3-3 \hat{\kappa}_4-2 \hat{\kappa}_6)\big]
\\
&+\frac{3}{25} F_2 \big[\Theta _{21} (6 \hat{\kappa}_1+15 \hat{\kappa}_2-16 \hat{\kappa}_3-3 \hat{\kappa}_4-2 \hat{\kappa}_6)+\Theta _{12} (6 \hat{\kappa}_1-\hat{\kappa}_2-3 \hat{\kappa}_5-2 \hat{\kappa}_6)\big]
%%%%%%%%
\\
\sum_i K^i_{221}F^i &= \frac{3}{25} (F^1+F^2 ) \big[\Theta _{12} (16 \hat{\kappa}_1-6 \hat{\kappa}_2-15 \hat{\kappa}_3+2 \hat{\kappa}_5+3 \hat{\kappa}_6)+\Theta _{21} (16 \hat{\kappa}_1-16 \hat{\kappa}_3-3 \hat{\kappa}_4+3 \hat{\kappa}_6)\big]
%%%%%%%%%%%%%%%
\\
\sum_i K^i_{211}F^i &= \frac{27}{25} (F^1+F^2) \big[\Theta _{12} (-2 \hat{\kappa}_1+2 \hat{\kappa}_2+\hat{\kappa}_5-\hat{\kappa}_6)+\Theta _{21} (-2 \hat{\kappa}_1+2 \hat{\kappa}_3+\hat{\kappa}_4-\hat{\kappa}_6)\big]
\end{align}
in which $\Theta_{12}=\theta(\eta -\eta_1) \theta(\eta_1-\eta_2)$ and $\Theta_{21}=\theta(\eta -\eta_2) \theta(\eta_2-\eta_1)$. Moreover $\sum_i K^i_{112}F^i$ and $\sum_i K^i_{212}F^i$ could be simply read off from $\sum_i K^i_{121}F^i$ and $\sum_i K^i_{221}F^i$, respectively, by the following replacements
 \begin{align}
\hat{\kappa}_2 \leftrightarrow \hat{\kappa}_3,\qquad \hat{\kappa}_4 \leftrightarrow \hat{\kappa}_5,\qquad F^1 \leftrightarrow F^2,\qquad F^3 \leftrightarrow F^4,\qquad \Theta_{12}\leftrightarrow \Theta_{21}
\end{align}

Note that the counterterms for the Euler equation, which are the ones with the first index 2, are all proportional to
\be
F_1+F_2 = k^2,\qquad \text{or}\qquad F_5 = k^2 \frac{(\k_1\cdot \k_2)^2}{k_1^2 k_2^2}
\ee
which in real space translate into $\nabla^2 \delta^2$ and $\nabla^2 (\d_i\d_j\phi)^2$, respectively. Hence, as expected they are Laplacians of locally observable scalar quantities and the corresponding force in the Euler equation is curl-free.

%%%%%%%%%%%%%%%%%%%%%%%%%%%%%%%%%%%%%%%%%%%%%%%%%%%
%%%%%%%%%%%%%%%%%%%%%%%%%%%%%%%%%%%%%%%%%%%%%%%%%%%
\begin{figure}
\begin{center}
    \begin{subfigure}[b]{0.35\textwidth}
        \centering
        \resizebox{\linewidth}{!}{
\begin{tikzpicture}[scale=.35]
\draw [line width= 2 pt](0,0) -- (6.06,6.06);
\draw [line width= 2 pt](3,3) -- (6,0);
\draw [line width= 2. pt](5.94,6.06) -- (10,2.0);

%\draw [line width= 2 pt](1.5,1.5) -- (3,0);
\draw [line width= 2 pt](10,2) -- (12,0);
\draw [line width= 2 pt](10.0,2.0) -- (8,0);

\draw [line width= .75 pt](6,6) -- (6,8.5);
\end{tikzpicture}
        }
        \caption{}
        \label{hardin1}
    \end{subfigure}
    \qquad
    \qquad
    \begin{subfigure}[b]{0.35\textwidth}
    \centering
        \resizebox{\linewidth}{!}{
            \begin{tikzpicture}[scale=0.35]

\draw [line width= 2 pt](0,0) -- (3.07,3.07);
\draw [line width= .75 pt](3,3.085) -- (6.,6.);
\draw [line width= 2 pt](3,3) -- (6,0);
\draw [line width= .75 pt](6,6) -- (10,2.085);

\draw [line width= 2. pt](1.5,1.5) -- (3,0);
\draw [line width= 2 pt](10,2) -- (12,0);
\draw [line width= 2 pt](10.07,2.07) -- (8,0);

\draw [line width= .75 pt](6,6) -- (6,8.5);
\end{tikzpicture}
        }
        \caption{}   
        \label{hardin2}
    \end{subfigure}
   
\caption{Hard ingoing lines combining into a soft line.}
\label{}
\end{center}
\end{figure}
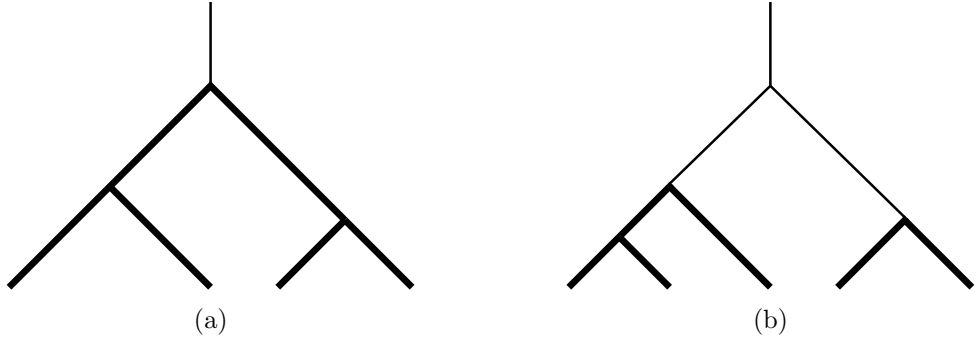

%%%%%%%%%%%%%%%%%%%%%%%%%%%%%%%%%%%%%%%%%%%%%%%%%%
\section{Softness properties of perturbation theory}\label{app:soft}

In this appendix we give a perturbative and iterative proof of the single softness of momentum $\pib(\k)$ and double softness of density contrast $\delta(\k)$ which result from evolution of short wavelength initial modes -- also known as decoupling. The proof is broken into several steps.

%%%%%%%%%%%%%%%%%%%%%%%%%%%%%%%%%%%%%%%%%%
\subsection{Only hard ingoing modes}\label{app:nosoft}

Let us first list the equations and constraints to be satisfied by $\delta$, $\bsb \nu = \nabla \times \pib$, and $\mu\equiv\nabla\cdot\pib$, namely \eqref{cont} and the divergence and the curl of \eqref{momentum}
\be\label{mu}
\dot\delta=-\nabla\cdot \pib,
\ee
\be\label{delta1}
\ddot\delta +\cH \dot\delta -\frac{3}{2}\cH^2 \delta = \d_i\d_j(\d_i\phi \d_j\phi)
-\frac{1}{2} \nabla^2(\d_j\phi \d_j\phi)
+\d_i\d_j(v^i\pi^j),
\ee
\be\label{nu}
\dot \nu^i +\cH \nu^i= -\vep^{ijk}\d_j[\d_m(\d_k\phi \d_m\phi)+\d_m(v^k\pi^m)]
\ee
where we used
\be\label{v}
\v = \frac{\pib}{1+\delta}
\ee
and
\be\label{phi}
\nabla^2 \phi =\frac{3}{2}\cH^2 \delta
\ee
to simplify the last two equations (and the subscript on $\v_\pi$ is dropped to avoid clutter). Suppose we start from a set of initial short wavelength (hard) modes. These modes evolve and combine according to the above equations to generate second and higher order perturbations. The softness properties follow from:

\begin{enumerate}

\item At any level when, as in figure \ref{hardin1}, hard modes combine into a soft $\delta$ through the vertices of \eqref{delta1}
\be
\delta_l(\eta) = \int^\eta_{\eta_0} d\eta' G_{\delta}(\eta,\eta')
[\d_i\d_j(\d_i\phi_s \d_j\phi_s)
-\frac{1}{2} \nabla^2(\d_j\phi_s \d_j\phi_s)
+\d_i\d_j(v_s^i\pi_s^j)]_{\eta'},
\ee
the result is automatically of second order in the outgoing momentum, i.e. $k_l^2$. (Here and in what follows we use the more compact subscripts $l$ and $s$ to denote long wavelength (soft) and short wavelength (hard), respectively.)

\item The same applies to $\nabla\cdot \pi_l$ because of the constraint \eqref{mu}, and \eqref{nu} ensures in a similar way that $\nabla\times \pib_l = \O(k_l^2)$. It follows that $\pib_l=\O(k_l)$.

\item If the hard modes combine into two long modes at a lower order which then evolve into one long $\delta_l$ (as in figure \ref{hardin2}), we have
\be\label{delta2}
\delta_l(\eta) = \int^\eta_{\eta_0} d\eta' G_{\delta}(\eta,\eta')
[\d_i\d_j(\d_i\phi_l \d_j\phi_l)
-\frac{1}{2} \nabla^2(\d_j\phi_l \d_j\phi_l)
+\d_i\d_j(v_l^i\pi_l^j)]_{\eta'}.
\ee
The softness of the final $\delta_l$ can be established by induction as follows.

\item If all long modes on the r.h.s. are the immediate result of combining short modes, we have from 1 and 2, and \eqref{v} and \eqref{phi}
\be\label{softness}
\d_i\phi_l =\O(k_l),\qquad v_l^i=\O(k_l^0),\qquad \pi^i_l = \O(k_l),
\ee
which ensures that $\delta_l$ on the left of \eqref{delta2} is at least $\O(k_l^3)$.\footnote{Since \eqref{v} is nonlinear and has no overall derivatives a priori one would expect $\v_l=\O(k_l^0)$. However, we will show in appendix \ref{app:v} that at lowest derivative order the (standard) perturbation theory results in $\v_l=\O(k_l)$.} This in turn ensures via \eqref{mu} that $\nabla\cdot \pib_l=\O(k_l^3)$ at this level, and one can similarly verify $\bsb \nu = \O(k_l^3)$. 

\item Therefore a stronger softness property than \eqref{softness} holds when the soft modes on the r.h.s. of \eqref{delta2} result from evolution of two soft modes that themselves evolved from hard modes, and the iteration can be continued ad infinitum. 

\end{enumerate}

%%%%%%%%%%%%%%%%%%%%%%%%%%%%%%%%%%%%%%%%%%%%%%%%%%%%%%%%%%%%%%%%
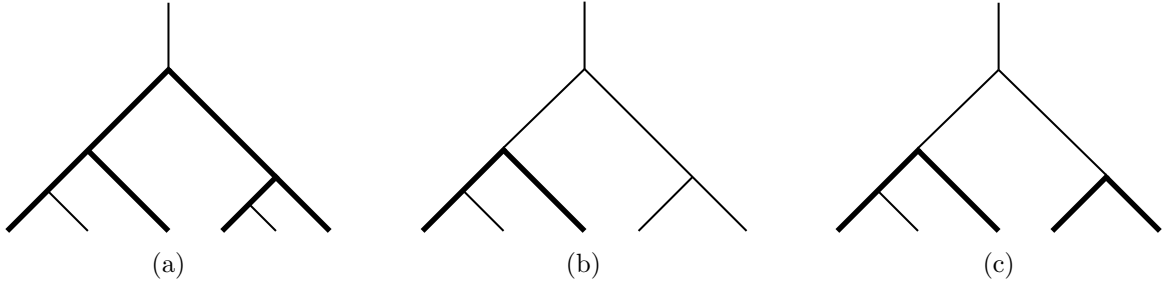
\begin{figure}
\begin{center}   
    \begin{subfigure}[b]{0.28\textwidth}
        \centering
        \resizebox{\linewidth}{!}{
\begin{tikzpicture}[scale=.35]
\draw [line width= 2 pt](0,0) -- (6.07,6.07);
\draw [line width= 2 pt](3,3) -- (6,0);
\draw [line width= .75 pt](1.5,1.5) -- (3,0);
\draw [line width= 2 pt](6,6) -- (12,0);
\draw [line width= 2 pt](10.07,2.07) -- (8,0);
\draw [line width= .75 pt](6,6) -- (6,8.5);
\draw [line width= .75 pt](10,0) -- (9,1);

\end{tikzpicture}
        }
        \caption{}
        \label{hsin1}
    \end{subfigure}
    \qquad
    \begin{subfigure}[b]{0.28\textwidth}
    \centering
        \resizebox{\linewidth}{!}{
            \begin{tikzpicture}[scale=0.35]
\draw [line width= 2 pt](0,0) -- (3.07,3.07);
\draw [line width= .75 pt](3,3.085) -- (6.,6.);
\draw [line width= 2 pt](3,3) -- (6,0);
\draw [line width= .75 pt](6,6) -- (12,0);
\draw [line width= .75 pt](1.5,1.5) -- (3,0);
\draw [line width= .75 pt](10.,2.0) -- (8,0);
\draw [line width= .75 pt](6,6) -- (6,8.5);

\end{tikzpicture}
        }
        \caption{}   
        \label{hsin2}
    \end{subfigure}
    \qquad
    \begin{subfigure}[b]{0.28\textwidth}
    \centering
        \resizebox{\linewidth}{!}{
            \begin{tikzpicture}[scale=0.35]

\draw [line width= 2 pt](0,0) -- (3.07,3.07);
\draw [line width= .75 pt](3,3.085) -- (6.,6.);
\draw [line width= 2 pt](3,3) -- (6,0);
\draw [line width= .75 pt](6,6) -- (10,2.085);

\draw [line width= .75 pt](1.5,1.5) -- (3,0);
\draw [line width= 2 pt](10,2) -- (12,0);
\draw [line width= 2 pt](10.07,2.07) -- (8,0);

\draw [line width= .75 pt](6,6) -- (6,8.5);
\end{tikzpicture}
        }
        \caption{}   
        \label{hsin3}
    \end{subfigure}

\caption{Hard and soft ingoing lines combining into a soft line.}
\label{hsin}
\end{center}
\end{figure}

%%%%%%%%%%%%%%%%%%%%%%%%%%%%%%%%%%%%%%%%%%%%%%%%%%%%%%%%
\subsection{Soft and hard ingoing modes}\label{app:wsoft}

In the presence of both soft and hard initial modes, double softness can be established by dividing the quadratic vertices of the perturbation theory into different categories. First, every vertex that results in a soft outgoing field either combines two short modes or two long modes, e.g.
\be
[\d\phi \d\phi]_l = \d\phi_s\d\phi_s +\d\phi_l\d\phi_l.
\ee
In the presence of initial soft modes we need to make a further distinction of whether a long mode is purely made of those initial long modes (indicated as $\d\phi_{l;l}$), or if it contains initial hard modes in addition to a (possibly empty) set of initial long modes (indicated as $\d\phi_{l;sl}$). Thus, the first vertex in \eqref{delta1} is decomposed schematically as
\be\label{phiphi}
\d\d[\d\phi_s\d\phi_s +\d\phi_{l;l}\d\phi_{l;sl} + \d\phi_{l;sl}\d\phi_{l;sl}].
\ee
Note that $\d\phi_{l;l}\d\phi_{l;l}$ is excluded since we are only interested in the case where there are some hard initial modes. Figure \ref{hsin} shows one example for any of the above three terms.

\begin{enumerate}

\item The first term in the above equation is similar what we considered in point 1 of the previous section, where the two external derivatives ensured double softness. However, there is an important difference: the short modes can now contain initial long modes (see figure \ref{hsin1}). One may therefore worry that $\d\phi_s$ may contain negative powers of $k_l$ due to the displacement of hard modes by the soft modes. That is, terms in perturbative evolution of the form $\v_{l;l}\cdot \nabla \delta_s$ where $\v_{l;l} = \O(k_l^{-1})$. These IR singularities indeed exist in any short wavelength quantity that evolves from initial hard and soft modes. They correspond to the motion of short modes with the bulk flow of the long modes and their presence is ensured by the equivalence principle. However, as shown more explicitly in appendix \ref{app:IRsing} whenever a collection of hard modes combine into a soft mode, adding any number of initial soft modes to such diagram does not lead to any negative power of $k_l$. The singular displacement of hard lines with momenta $\{\q_i\}$ by a soft line of momentum $\k_1$ that are proportional to $\k_1\cdot\q_i/k_1^2$ add up to $\k_1\cdot \k_{\rm total}/k_1^2 -1 =\O(k_l^0)$ at any moment. Hence, the first term in \eqref{phiphi} leads to a double soft outgoing $\delta_l$ (and as before a single soft $\pib_l$).

\item The second and third terms can also be shown to lead to at least double soft results using induction. First suppose there is no initial long modes in $\d\phi_{l;sl}$. Then by the argument of the last section $\d\phi_{l;s}=\O(k_l)$. And $\d\phi_{l;l} = \O(k_l^{-1})$ from \eqref{phi}. So the resulting outgoing $\delta_{l;sl} = \O(k_l^2)$ because of the overall derivatives. One can then continue the induction provided that the same argument applies to the $\d\d(v\pi)$ vertex of \eqref{delta1}.

\item Decomposing $v\pi$ into short and long wavelength contributions we get
\be\label{vpi}
[v\pi]_l = v_s\pi_s +v_{l;sl}\pi_{l;l}+ v_{l;sl}\pi_{l;sl}+v_{l;l}\pi_{l;sl}.
\ee
Assuming that up to some order in perturbations 
\be\label{softness2}
\pi_{l;sl}=\O(k_l),\qquad \delta_{l;sl}=\O(k_l^2),\qquad v_{l;sl}=\O(k_l^0),
\ee
and using the fact that $v_{l;l},\pi_{l;l}=\O(k_l^{-1})$ and $\delta_{l;l}=\O(k_l^0)$, we see that the last two terms on the r.h.s. of \eqref{vpi} preserve these properties (when considering the vertex $\d\d( v\pi)$). The second term clearly violates them and the first term does so too: expanding
\be
\begin{split}
v_s\pi_s = & v_s [\delta_s v_s + (1+\delta)_l v_s +\delta_s v_l]\\[10pt]
= & v_s [\delta_s v_s + (1+\delta)_l v_s +\delta_s v_{l;sl}+\delta_s v_{l;l}],
\end{split}
\ee
the last term is seen to be $\O(k_l^{-1})$, leading to $\O(k_l)$ contribution to $\delta_l$ from the vertex $\d\d(v_s\pi_s)$. However, the sum of the two dangerous terms preserves the condition \eqref{softness2}. Substituting 
\be
\pi_{l;l} = (1+\delta)_{l;l} v_{l;l}
\ee
in the second term on the r.h.s. of \eqref{vpi} and using the fact that $(1+\delta)_{l;sl} = \delta_{l;sl}$ (since the subscript $s$ indicates a nonzero set of initial short modes) we obtain for the sum of potentially dangerous terms
\be
\begin{split}
v_{l;l}[(1+\delta)_{l;l}v_{l;sl}+\delta_s v_s]
&=v_{l;l}\left\{[(1+\delta)v]_{l;sl} - (1+\delta)_{l;sl} (v_{l;l}+v_{l;sl})\right\}\\[10pt]
&=v_{l;l}[\pi_{l;sl} -\delta_{l;sl} (v_{l;l}+v_{l;sl})] = \O(k_l^0).
\end{split}
\ee
Therefore the sum of all vertices preserves the condition \eqref{softness2}, and double softness follows by induction.
\end{enumerate}

Note finally that all of the above arguments generalize to the case where counterterms of the form $\d_j\tau^{ij}$, with $\tau^{ij}$ a locally observable tensor which by definition can never have an IR singularity (corresponding to a negative scaling with $k_l$), are added to the Euler equation \eqref{momentum}. This seems to be the most general way of modifying the equations of motions that is still compatible with locality and momentum conservation. Intuitively, the short scale dynamics can generate an effective force on a volume element of the fluid only due to its momentum flux into that volume. 

%%%%%%%%%%%%%%%%%%%%%%%%%%%%%%%%%%%%%%%%%%%%%%%%%%%%%%%%%%%%
\subsection{Velocity field}\label{app:v}

The final point to be discussed is the softness property of the mass weighted velocity field defined as $\v=\pi/(1+\delta)$. Using mass conservation we get
\be\label{theta1}
\theta =\nabla\cdot \v= -\dot \delta -\nabla\cdot(\delta \v).
\ee
Using this equation in the Euler equation leads to
\be\label{delta4}
\ddot\delta+\cH\dot\delta-\frac{3}{2}\cH^2 \delta = \cH(\d_\tau +\cH)\d_i(\delta \d_i\phi_v)
+\cH^2\nabla^2(\d_i\phi_v\d_i\phi_v),
\ee
where we neglected vorticity and set $\v =-\cH \nabla \phi_v$. From these equations it follows that double-softness of a soft outgoing $\delta$ evolved purely from initial hard modes implies double softness of $\theta$. To show this one needs to argue that the second term in \eqref{theta1} which naively is only single soft is actually double-soft when the perturbative solution for $\delta$ and $\theta$ are substituted. This follows from the fact that the second vertex on the r.h.s. of \eqref{delta4} is automatically double soft and if $\delta$ is to be double soft the first term on the r.h.s. must also be double soft by itself.\footnote{One should also consider the case when the hard modes combine into two soft modes which later combine into a final $\delta$. In this case too the second term on the r.h.s. of \eqref{delta4}
\be
\nabla^2(v_{l;s} v_{l;s})
\ee
is double soft because at least $v_{l;s}=\O(k_l^0)$.}

If we use $\delta$ and $\v$ as primary variables -- as we did in our systematic renormalization -- and add counter-terms proportionally to SPT cutoff dependence, the resulting renormalized velocity inherits both curl-freedom and single-softness from SPT kernels. This renormalized velocity of course differs from $\v_\pi$. On the other hand, if we use $\v_\pi$ and add counterterms as $\d_j\tau^{ij}$ to the momentum equation \eqref{momentum}, they modify \eqref{delta4} as $\d_i[(1+\delta)^{-1}\d_j\tau^{ij}]$. Expanding this leads to vertices which are not explicitly second derivative. For instance, taking $\tau^{ij}=\delta^{ij}(\d_m\d_n\phi)^2$ we get an associated cubic vertex
\be
-\d_i[\delta\ \d_i(\d_m\d_n\phi)^2].
\ee
Taking all ingoing modes to be hard, this vertex gives rise to an outgoing $\delta$ which is only single soft. In this formulation, the full result becomes double soft (as it must according to the arguments of sections \ref{app:nosoft} and \ref{app:wsoft}) because the above single soft term cancels with the one arising from evolving a second order solution coming from the leading part of the counterterm $\nabla^2 (\d_m\d_n\phi)^2$ to third order via the vertex $\d_i(\delta v^i)$. Hence, the double softness of $\delta$ does not imply that of $\theta_\pi$ anymore because the second term on the r.h.s. of \eqref{theta1} is generically only single soft after the addition of counterterms. Note that one can redefine velocity to absorb these single soft terms, but the new velocity is no longer given by $\pib/(1+\delta)$.

%%%%%%%%%%%%%%%%%%%%%%%%%%%%%%%%%%%%%%%%%%%%%%%%%%%
%%%%%%%%%%%%%%%%%%%%%%%%%%%%%%%%%%%%%%%%%%%%%%%%%
\section{Cancellation of IR singularities}\label{app:IRsing}

Consider the following time evolution diagram with ingoing momenta $\{\k_1,\q_1,\q_2\}$ and outgoing momentum $\k$, and such that $k_1\sim k \ll q_{1,2}$.
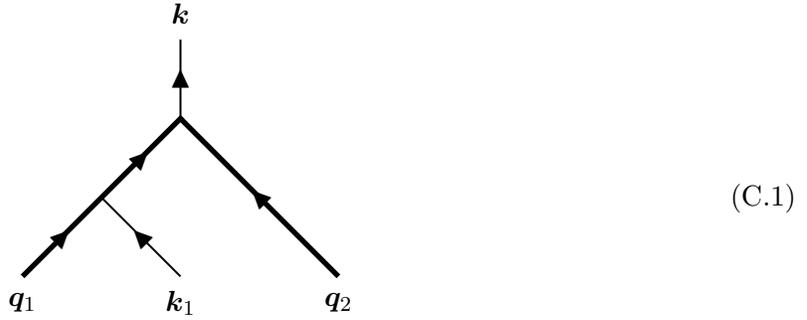
\be\label{delta3}
\raisebox{-45 pt}{
\begin{tikzpicture}[scale= .7]
\draw [line width= 2 pt](0,0) -- (3.035,3.035);
\draw [line width= 2 pt](3,3) -- (6,0);

\draw [line width= .75 pt](3,3) -- (3,4.5);
\draw [line width= .75 pt](3,0) -- (1.5,1.5);

\draw (.75,.75) node [rotate=45] {$\blacktriangleright$};
\draw (2.25,2.25) node [rotate=45] {$\blacktriangleright$};
\draw (2.25,.75) node [rotate=-45] {$\blacktriangleleft$};
\draw (3,3.75) node [rotate=90] {$\blacktriangleright$};
\draw (4.5,1.5) node [rotate=-45] {$\blacktriangleleft$};

\draw (0,-.5) node  {$\Q_1$};
\draw (3,-.5) node  {$\K_1$};
\draw (6,-.5) node  {$\Q_2$};
\draw (3,5) node  {$\K$};

\end{tikzpicture}}
\ee
The potential IR singularity arises in this diagram from the vertex $\v\cdot \nabla \Psi_a$, where $\v$ realizes the soft $\k_1$ mode and $\Psi_a$ a hard one, leading to $\k_1\cdot \q_i/k_1^2$. Focusing only on this vertex we have for the outgoing field
\be
\begin{split}
\{\Psi_a(\k,\eta)\}_{\rm IR}=\int^\eta_{\eta_0}d\eta_1 g_{ab}(\eta,\eta_1)&\int^{\eta_1}_{\eta_0} d\eta_2
[\gamma_{bcd}(\k,\q_1+\k_1,\q_2)g_{ce}(\eta_1,\eta_2)
\frac{\k_1\cdot \q_1}{k_1^2} \\[10pt]
&\Psi_2^{(1)}(\k_1,\eta_2)\Psi_e^{(1)}(\q_1,\eta_2)
\Psi_d^{(1)}(\q_2,\eta_1)+ \q_1\leftrightarrow \q_2].
\end{split}
\ee
Now using
\be
\Psi_d(\q_2,\eta_1)=g_{df}(\eta_1,\eta_2)\ \Psi_f(\q_2,\eta_2),\\[10pt]
\ee
and the fact that $\gamma_{bcd}(\k,\q_1+\k_1,\q_2)=\gamma_{bcd}(\k,\q_1,\q_2)+\O(k_1)$ in the limit $k_1\ll q_i$, we obtain
\be
\begin{split}
\{\Psi_a(\k,\eta)\}_{\rm IR}=&\int^\eta_{\eta_0}d\eta_1 g_{ab}(\eta,\eta_1)\int^{\eta_1}_{\eta_0} d\eta_2
\gamma_{bcd}(\k,\q_1,\q_2)g_{ce}(\eta_1,\eta_2)g_{df}(\eta_1,\eta_2)\\[10pt]
&\left[\frac{\k_1\cdot \q_1}{k_1^2}+\frac{\k_1\cdot \q_2}{k_1^2}\right]
\Psi_2^{(1)}(\k_1,\eta_2)\Psi_e^{(1)}(\q_1,\eta_2)
\Psi_f^{(1)}(\q_2,\eta_2).
\end{split}
\ee
The expression in the square brackets is $\k_1\cdot(\k-\k_1)/k_1^2$ which is $\O(k^0)$, thus the potential IR singularities coming from the $\v\cdot \nabla \Psi_a$ vertex cancel in this diagram when we sum over the two possible attachments of the soft line. The full diagram will therefore have the same softness property as the one without attaching the soft line. 

The above cancellation arises from the fact that the displacement produced by the long $\k_1$ mode is universal, and hence when the individual hard momenta add up to a soft momentum $\k$, the displacement terms also add up as effectively moving a long wavelength mode. Note that the cancellation happens independently at any time-slice. Next we show this to hold more generally whenever one or several ingoing soft modes are added to a diagram that combines a set of hard modes into a soft mode. 

Let us focus again on attaching one soft line via the $\v\cdot \nabla$ vertex at a specific time $\eta_1$, with 
\be
\v=-\cH\frac{\k_1}{k_1^2} \Psi_2^{(m)}(\k_1,\eta_1)
\ee
an order $m$ soft velocity, and freeze the rest of the diagram. Cutting this frozen diagram at time $\eta_1$ and stripping off the soft $\k_1$ line we have a set of fields $\{\Psi_{a_i}^{(n_i)}(\q_i,\eta_1)\}$ whose contribution to the outgoing field is
\be
A_{a a_1\cdots}(\eta;\eta_1,\{\q_i\}) \prod_i\Psi_{a_i}^{(n_i)}(\q_i,\eta_1),
\ee
where $A_{a\cdots}$ is a function that contains all momentum dependences arising from the vertices after $\eta_1$ and all propagators. Now add the soft line and sum over all possible attachments of $\v\cdot \nabla$ vertex at $\eta_1$
\be
\sum_i A_{a a_1\cdots}(\eta,\eta_1,\{\q_1,\cdots,\q_i+\k_1,\cdots\}) \frac{\k_1\cdot \q_i}{k_1^2}
\Psi_2^{(m)}(\k_1,\eta_1)\prod_i\Psi_{a_i}^{(n_i)}(\q_i,\eta_1),
\ee
and use the fact that $A_{a \cdots}(\eta,\eta_1,\{\q_1,\cdots,\q_i+\k_1,\cdots\}$ is regular when $k_1\ll q_i$ to get
\be
A_{a a_1\cdots}(\eta,\eta_1,\{\q_1,\cdots,\q_i,\cdots\}) \sum_i\frac{\k_1\cdot \q_i}{k_1^2}
\Psi_2^{(m)}(\k_1,\eta_1)\prod_i\Psi_{a_i}^{(n_i)}(\q_i,\eta_1) +\O (k_1^0).
\ee
The sum gives $\k_1\cdot(\k-\k_1)/k_1^2 = \O(k_l^0)$. Hence, the drifts caused by a long wavelength mode add up at any moment to the overall drift of the total (stripped) diagram whose momentum is $\k-\k_1$. This proves that if we start from a composite operator such as $\d\phi_s\d\phi_s$, which combines hard modes into a soft outgoing line, and perturbatively add soft modes they will not introduce any negative power of $k_l$.

%%%%%%%%%%%%%%%%%%%%%%%%%%%%%%%%%%%%%%%%%%%%%%%%%%
 
\section{Equations in Fourier space}\label{a:f}

In this appendix we provide explicit formulae for the equations of motion of momentum and velocity in Fourier space. 

The equations of motion using momentum are
\ba
\partial_{\tau} \dl+\ml&=&0\,,\\
\partial_{\tau}\ml+ \cH \ml+\frac{3}{2}\cH^{2}\Omega_{m}\dl&=& \xi^{\dl\dl}+\xi^{\ml\ml\dl}+\xi^{\ml\nl\dl}+\xi^{\nl\nl\dl}+\partial_{i}\partial_{j}\tau^{ij}\,\\
\partial_{\tau}\nl^{i}+ \cH \nl^{i}&=& \vec \gamma^{\ml\ml\dl}+ \vec \gamma^{\ml\nl\dl}+ \vec \gamma^{\nl\nl\dl}+\epsilon^{ijk}\partial_{j}\partial_{m}\tau^{mk}\,.
\ea
Several definitions are in order. $\xi^{\dl\dl}$ comes from the term $\delta \partial_{i} \phi$  
\ba
\xi^{\dl\dl}&\equiv & -\frac{3}{2} \cH^{2}\Omega_{m} \intd{\vec q}\alpha(\vec q, \vec k-\vec q) \dl(\vec q) \dl(\vec k-\vec q)\nonumber\\
&\rightarrow& -\frac{3}{2} \cH^{2}\Omega_{m} \intd{\vec q} \alpha^{(s)}(\vec q, \vec k-\vec q)  \dl(\vec q) \dl(\vec k-\vec q)\,,
\ea
where $ \alpha $ was given in \eqref{ab} and the arrow refers to symmetrization\footnote{\label{f1}The reason why we drop the antisymmetric term is the following. For any anti-symmetric function $A(\vec k,\vec q)=-A(\vec q,\vec k)$ one has
\be
\intq A(\vec q,\vec k-\vec q)=-\intq A(\vec q,\vec k-\vec q)\,,
\ee
as it can be easily proven. This equation has two solutions: zero and infinity. In applications to perturbation theory the integral is often divergent. We regulate the divergences in a way that respect the anti-symmetry so that the regularized terms vanish. E.g.~a divergence in a integral like
\be
\intd{\vec q} \alpha(\vec q,\vec k-\vec q) \delta(\vec q)\delta(\vec k-\vec q)\,,
\ee
is regulated by smoothing each $\delta(k)$ on some cutoff scale $k_{\rm cut off}$.} 
\ba
\alpha^{(s)}(\vec k, \vec q)&\equiv& \frac{1}{2} \left[\alpha(\vec k,\vec q)\pm \alpha(\vec q,\vec k)\right]= \frac{(\vec k+\vec q)}{2} \cdot\left(\frac{\vec q}{q^{2}}+\frac{\vec k}{k^{2}}\right)\,.
\ea
The other interactions come from $\partial_{i} \left(\pl^{i}\pl^{k}/\rho_{l}\right)$ and are as follows.
\ba
\xi^{\mu\mu\delta}(\vec k)&=&-\sum_{n=0}^{\infty} \intd{\vec q_{1},\vec q_{2}} \, \frac{\vec k \cdot \vec q_{2}\vec k \cdot \vec q_{12}}{q_{2}^{2} q_{12}^{2}} \mu(\vec q_{2})\mu(\vec q_{12})\, \delta^{n}(\vec k-\vec q_{1}) (-1)^{n}\,,\non \\
\xi^{\mu\nu\delta}(\vec k)&=&\sum_{n=0}^{\infty} \intd{\vec q_{1},\vec q_{2}} \, \frac{\vec k \cdot \vec q_{2}\times \vec \nu(\vec q_{2})\,\vec k \cdot \vec q_{12} \mu(\vec q_{12})-\vec k \cdot \vec q_{12}\times \vec \nu(\vec q_{12})\,\vec k \cdot \vec q_{2}  \mu(\vec q_{2})}{q_{2}^{2} q_{12}^{2}} \delta^{n}(\vec k-\vec q_{1})(-1)^{n}\,, \non \\
\xi^{\nu\nu\delta}(\vec k)&=&-\sum_{n=0}^{\infty} \intd{\vec q_{1},\vec q_{2}} \, \frac{\vec k \cdot \vec q_{2}\times \vec \nu(\vec q_{2}) \vec k \cdot \vec q_{12}\times \vec \nu(\vec q_{12})}{q_{2}^{2} q_{12}^{2}} \, \delta^{n}(\vec k-\vec q_{1})(-1)^{n}\,,\non \\
\vec \gamma^{\mu\mu\delta}(\vec k)&=&-\sum_{n=0}^{\infty} \intd{\vec q_{1},\vec q_{2}} \, \frac{\vec k \times \vec q_{2}\vec k \cdot \vec q_{12}}{q_{2}^{2} q_{12}^{2}} \mu(\vec q_{2})\mu(\vec q_{12})\, \delta^{n}(\vec k-\vec q_{1}) (-1)^{n}\,,\non \\
\vec \gamma^{\mu\nu\delta}(\vec k)&=&\sum_{n=0}^{\infty} \intd{\vec q_{1},\vec q_{2}} \, \frac{\vec k \times \left[ \vec q_{2}\times \vec \nu(\vec q_{2}) \right] \,\vec k \cdot \vec q_{12} \mu(\vec q_{12})-\vec k \times \left[ \vec q_{12}\times \vec \nu(\vec q_{12}) \right]\,\vec k \cdot \vec q_{2}  \mu(\vec q_{2})}{q_{2}^{2} q_{12}^{2}} \delta^{n}(\vec k-\vec q_{1})(-1)^{n}\,, \non \\
\vec \gamma^{\nu\nu\delta}(\vec k)&=&-\sum_{n=0}^{\infty} \intd{\vec q_{1},\vec q_{2}} \, \frac{\vec k \times \left[ \vec q_{2}\times \vec \nu(\vec q_{2}) \right]  \vec k \times \left[ \vec q_{12}\times \vec \nu(\vec q_{12})\right]}{q_{2}^{2} q_{12}^{2}} \, \delta^{n}(\vec k-\vec q_{1})(-1)^{n}\,,\nonumber
\ea
where $\vec q_{12}\equiv \vec q_{1}-\vec q_{2}$, $\delta^{n}(\vec k)$ is the Fourier transform of $\delta(\vec x)^{n}$ and we used the vector identity 
\be
\vec a\times \vec b \times \vec c=\vec b \,\vec a \cdot \vec c-\vec c\, \vec a \cdot \vec b
\ee
Moving on to the velocity equations, we start with the Fourier-space decomposition
\be
v^{i}(\vec k)=i \frac{\epsilon^{ijj'}k_{j} w_{j'}(\vec k)}{k^{2}} -i\frac{k^{i}}{k^{2}}\theta(\vec k)\,.
\ee
The equations of motion in terms of velocity gradient and curl in Fourier space, supplemented by the EFT corrections, are then given by
\ba
\partial_{\tau} \delta+ \theta &=& \intq \left[-\alpha(\vec q,\vec k-\vec q) \theta(\vec q) + \va(\vec q,\vec k-\vec q) \cdot\vec w(\vec q) \right] \delta(\vec k-\vec q)\,,\\
\partial_{\tau} \theta+ \cH \theta+\frac{3}{2} \cH^{2}\Omega_{m} \delta&=& -\intq \left[\beta(\vec q,\vec k-\vec q) \theta(\vec q)\theta(\vec k-\vec q)+ \vb(\vec q,\vec k-\vec q) \cdot\vec w(\vec k-\vec q) \theta(\vec q)+\right.\non \\ &&\quad \left.  \vec w (\vec q)\mb (\vec q,\vec k-\vec q) \vec w(\vec k-\vec q)\right]+\partial_{i}\left( \frac{\partial_{j}\tau^{ij} }{1+\delta}\right) \\
\partial_{\tau} \vec w +\cH \vec w&=&\intq \mg_{ij}(\vec q,\vec k-\vec q) w^{j}(\vec k-\vec q) \theta(\vec q)+\va(\vec q,\vec k-\vec q) \vec w(\vec q)\cdot  \vec w(\vec k-\vec q)\nonumber\\
&&\quad +\vg_{ijl}(\vec q,\vec k-\vec q) w^{l}(\vec k-\vec q) w^{j}(\vec q)+\epsilon^{ijk}\partial_{j}\left( \frac{\partial_{l}\tau^{kl} }{1+\delta}\right) \,,
\ea
where $ \alpha $ and $ \beta $ were given in \eqref{ab} and\footnote{We used the magic vector identity
\be
\vec \partial\times \left(v^{k}\partial_{k} \vec v\right)=- \vec \partial \times \left[\vec v \times \left(\vec \partial\cdot \vec v\right)\right]\,.
\ee}
\ba
%\alpha (\vec q_{1},\vec q_{2})&\equiv&\frac{(\vec q_{1}+\vec q_{2})\cdot \vec q_{1}}{q_{1}^{2}} \,, \quad 
\va  (\vec q_{1},\vec q_{2})&=& \frac{\vec q_{2} \times \vec q_{1}}{q_{1}^{2}}\,,
%\beta(\vec q_{1},\vec q_{2})&\equiv &\frac{ \left(\vec q_{1} +\vec q_{2}\right)^{2}\vec q_{1} \cdot \vec q_{2} }{2 q_{1}^{2}q_{2}^{2}}\,,
\quad \vb(\vec q_{1},\vec q_{2})=\frac{\vec q_{1}\times \vec q_{2}}{q_{2}^{2}}\,,\quad \nonumber
\mb_{ij}(\vec q_{1},\vec q_{2})\equiv \frac{ \left(\vec q_{1}\times \vec q_{2}\right)_{i}\left(\vec q_{1}\times \vec q_{2}\right)_{j}}{q_{1}^{2}q_{2}^{2}}\,,\\
\mg_{ij}(\vec q_{1},\vec q_{2})&\equiv& \frac{\vec q_{1,i} \left(\vec q_{1}+\vec q_{2}\right)_{j}}{q_{1}^{2}}-\frac{\left(\vec q_{1}+\vec q_{2}\right)\cdot q_{1}}{q_{1}^{2} } \delta_{ij}\,,\quad \vg_{ijl}(\vec q_{1},\vec q_{2})\equiv \frac{\epsilon^{il'l} (\vec q_{1}+\vec q_{2})_{l'} q_{1}^{j}}{q_{1}^{2}}\,.
\ea

%%%%%%%%%%%%%%%%%%%%%%%%%%%%%%%%%%%%%%%%%%%%%%%%%%%%%%%%%%%%%%%%%%%%%
\section{Glossary}\label{app:glos}

In this paper, we have extensively used the same nomenclature as in quantum and statistical field theory to stress the similarity with the problem at hand and to make the discussion more intuitive. However, when needed we have adapted and refined some of the definitions. For the convenience of the reader we have collected below those technical terms used in this paper whose meaning differs in some important or subtle way from the common usage. 

\vspace{1cm}

\renewcommand{\arraystretch}{1.5}
\noindent
\begin{tabular}{l  p{10cm}}
%{\bf Propagator }& \\ 
%{\bf Correlator} & \\ 
{\bf 1-Particle Reducible (1PR)}& A loop diagram for a correlation function that can be divided into two disconnected parts by cutting an internal line that is not immediately connected to a contraction or an external line, e.g. \eqref{B1PR}\\
{\bf 1-Particle Irreducible }& Any loop diagram that is not 1PR, e.g. \eqref{P13}\\
{\bf Time-evolution diagram} & It shows the evolution of a set of initial fields into one higher order final field at a later time  \eqref{t1PIa}\\
{\bf Hard line} &  In a diagram, a line representing the time evolution of perturbation $ \delta(k) $ with a large momentum $ k\rightarrow \infty $, e.g. thick lines in \eqref{t1PIa} \\
{\bf Soft line} & In a diagram, a line representing the time evolution of perturbation $ \delta(k) $ with a small momentum, e.g. thin lines in \eqref{t1PIa}   \\
{\bf Amputated diagram}& A time evolution diagram in which the soft lines are immediately connected to the hard lines, e.g. \eqref{P1PI} \\
{\bf Non-stochastic} & The counterterm of a time-evolution diagram in which all of hard lines are paired and contracted with each other, e.g. \eqref{Psi3}\\
{\bf Stochastic} & The counterterm of a time-evolution diagram in which one or more hard ingoing lines remain unpaired, e.g. \eqref{Psi2}\\
{\bf Double softness} & A field is double soft iff it scales as $ k^{2} $ as $ k\rightarrow 0 $, e.g. $ \delta $ and $ \nabla \pi $ \\
{\bf Systematic} & Renormalization is systematic if new counterterm are needed only for 1PI diagrams, while divergences in 1PR diagrams are already canceled by lower order counterterms
\end{tabular}

%%%%%%%%%%%%%%%%%%%%%%%%%%%%%%%%%%%%%%%%%%%%%%%%%%%%%%%

\bibliographystyle{utphys}
\bibliography{sys_ren}
\end{document}